\documentclass[12pt,preprint]{aastex}

\shorttitle{Constraints on Planet Populations}
\shortauthors{Nielsen et al.}

\begin{document}


\title{Constraints on Extrasolar Planet Populations from VLT NACO/SDI and MMT SDI and Direct Adaptive Optics Imaging Surveys: Giant Planets are Rare at Large Separations}


\author{Eric L. Nielsen\altaffilmark{1}, Laird M. Close, and Beth A. Biller}
\affil{Steward Observatory, University of Arizona, Tucson, AZ 85721}
\email{enielsen@as.arizona.edu}

\and

\author{Elena Masciadri}
\affil{INAF-Osservatorio Astrofisico di Arcetri, Italy}

\and

\author{Rainer Lenzen}
\affil{Max-Planck-Institut f\"{u}r Astronomie, K\"{o}nigstuhl 17, 69117 Heidelberg, Germany}




\altaffiltext{1}{Michelson Fellow}


\begin{abstract}
We examine the implications for the distribution of extrasolar planets based 
on the null results from two of the largest direct imaging surveys published 
to date.  Combining the measured 
contrast curves from 22 of the stars observed with the VLT NACO adaptive 
optics system by \citet{elena}, and 48 of the stars observed with the VLT 
NACO SDI and MMT SDI 
devices by \citet{sdifinal} (for a total of 60 unique stars: the median star 
for our survey is a 30 Myr K2 star at 25 pc), we consider what distributions 
of planet masses 
and semi-major axes can be ruled out by these data, based on Monte Carlo 
simulations of planet populations.  We can set the following 
upper limit with 95\% confidence: the fraction of stars with planets with 
semi-major axis 
between 20 and 100 AU, and mass above 4 M$_{Jup}$, is 20\% or less.  Also, 
with a distribution of planet mass of 
$\frac{dN}{dM} \propto M^{-1.16}$ in the range of 0.5-13 M$_{Jup}$, 
we can rule out a power-law distribution for semi-major axis 
($\frac{dN}{da} \propto a^{\alpha}$) with index 
0 and upper cut-off of 18 AU, and index -0.5 with an upper cut-off of 48 AU.  
For the distribution suggested by \citet{cumming}, a power-law 
of index -0.61, we can place an upper limit of 75 AU on 
the semi-major axis distribution.  
At the 68\% confidence level, these upper limits state that fewer than 
8\% of stars have a planet of mass $>$4 M$_{Jup}$ between 20 and 100 AU, and a 
power-law distribution for semi-major axis with index 0, -0.5, and -0.61 
cannot have giant planets beyond 12, 23, and 29 AU, respectively.  In 
general, we find that even null results from direct imaging surveys are very 
powerful in constraining the distributions of giant planets (0.5-13 M$_{Jup}$) 
at large separations, but more work needs to be done to 
close the gap between planets that can be detected by direct imaging, and 
those to which the radial velocity method is sensitive.
\end{abstract}





\section{Introduction}

There are currently well over 200 known extrasolar planets, the bulk of which 
were discovered by radial velocity surveys (e.g. \citet{rvref}).  While 
this field has initially been dominated by the study of the relatively 
easy-to-find Hot Jupiters (planets with orbital periods of order days), over 
the past several years there has been an increasing amount of data describing 
planets in larger orbits.  In particular, \citet{fv05} compared radial 
velocity target stars with known planets to stars that had been monitored 
but did not show signs of planets; they concluded that about 5\% of stars 
had planets of mass greater than 1.6$M_{Jup}$, in orbits shorter than 4 years 
(within 2.5 AU).  Additionally, they determined that planet fraction 
increased with the host star's metal abundance.  \citet{rvref} have 
also considered the distributions 
of semi-major axis and planet mass of known radial velocity planets, and 
found that both distributions are well-fit by power laws.  \citet{cumming} 
have examined the biases of the radial velocity technique, and found 
that the semi-major axis distribution found by \citet{rvref}, 
$\frac{dN}{dP} \propto P^{-1}$, should be modified in light of the decreasing 
sensitivity of the radial velocity method with orbital distance, and suggest a
power law index of -0.74 for period, instead (for solar-like stars, this 
corresponds to a power law distribution for semi-major axis where
$\frac{dN}{da} \propto a^{-0.61}$).

A careful consideration of sensitivity of microlensing observations to 
planets by \citet{microlens1} suggests that for certain lensing 
geometries, at projected separations of $\sim$1-4 AU, the lower limit for 
the frequency of Neptune-mass planets is 16\%, making low-mass planets more 
common than giant planets in the inner solar system (though we note that the 
range of separations probed by \citet{microlens1} and \citet{fv05} do not 
precisely overlap, and the target star samples are not uniform between the 
two surveys).  Additionally, \citet{microlens2} found that from 
existing microlensing data, a third or less of M dwarfs in the galactic 
bulge have 1 M$_{Jup}$ planets in orbits between 1.5 and 4 AU, and 
$\leq$45\% of M dwarfs have planets between 1 and 7 AU of mass  3 M$_{Jup}$.

One outstanding question is how the abundance of planets varies as one 
considers planets in longer orbits.  \citet{raymond06} has studied the 
dynamics of terrestrial planet formation in systems with giant planets, and 
found from numerical simulations that giant planets impede the formation of 
earth-like planets when the giant planet orbits within 2.5 AU, and that water 
delivery to a terrestrial planet is only possible in significant amounts 
when the giant planet is beyond 3.5 AU.  The full extent to which giant 
planets impede (or encourage) water-rich terrestrial 
planet formation is still unknown.  A greater understanding of the 
distribution of giant planets is a precursor to investigating the conditions 
under which habitable terrestrial planets form and evolve.

The global distribution of giant planets has also been considered from the 
theoretical direction.  \citet{idalin} have produced 
distributions of planets forming in disks by core accretion, showing a 
continuation of a power law from the radial velocity regime (within 2.5 AU) 
for giant planets, out to about 10 AU, then trailing off at larger radii.  
It is possible that the lack of outer planets in these simulations may be 
due (at least in part) to the fact that these models do not consider the 
effects of planet-planet scattering after planets are formed, or it may 
simply be a function of the initial conditions of the simulation.  In 
order to constrain such models it is necessary to measure the distribution 
of giant planets in longer orbits, so as to fully sample parameter space.

With the advent of adaptive optics (AO) systems on large ($\sim$8m) 
telescopes, the ability to detect and characterize planets by directly imaging 
the companion is becoming increasingly viable.  Already planetary mass 
companions (in most cases $\sim$13 M$_{Jup}$ at 40-300 AU, or even lower mass 
objects with brown dwarf hosts) have been detected in certain favorable 
circumstances (e.g. 
companions to 2MASS1207: \citet{2mass1207}, AB Pic: \citet{abpic}, Oph 
1622: \citet{oph1622b}, \citet{oph1622c}, \citet{oph1622a}, CHXR 73: 
\citet{chxr73}, and DH Tau: \citet{dhtau}), and numerous surveys are underway 
for planets around nearby, young stars (since a self-luminous planet 
is brightest at young ages).  While the paucity of traditional planets 
(that is, planets $<$13 M$_{Jup}$ and $<$40 AU orbiting a star) detected by 
this method has been disappointing, in this paper we consider how even a 
null result from these direct imaging surveys can be used to set constraints 
on the population of giant planets.  As the sensitivity of radial velocity 
surveys to planets at larger separations decreases (due both to the smaller 
radial velocity signal, and the much longer orbital period requiring a longer 
time baseline of observations to adequately constrain the orbital 
parameters), at orbits wider than 10 AU only direct imaging is efficient at 
characterizing the extrasolar planet population.


\citet{janson} used VLT SDI data (part of the data considered in this work) 
for the known planet host star $\epsilon$ Eridani, to search for the radial 
velocity planet, given its predicted position from the astrometric orbit 
of \citet{benedict}.  Though upper limits were found for the planet of 
$M_{H}\sim$19, the predicted flux of the planet could be up to 10 magnitudes 
fainter, given the likely age of the system of 800 Myr.  While this is 
young compared to the rest of the radial velocity planet host stars, it is 
quite old by the standards of direct imaging planet searches, so the inability 
to detect this planet's flux is unsurprising.  Previous work has been done by 
\citet{kasper} to study the region of parameter space unprobed by the radial 
velocity method, large orbital separations, by observing 22 young, nearby 
stars in the L-band from the VLT.  The null result from this survey was used 
to set constraints on combinations of power-law index and upper cut-off for 
the distribution of the observed separation (not semi-major axis) of 
extrasolar planets.

\citet{elena} conducted a survey of 28 young, nearby stars, with a null result 
for planets.  They found that their observations were 
sufficient to detect a 5 M$_{Jup}$ planet at projected separations greater 
than 14 AU around 14 of their target stars, and above 65 AU for all 28 stars.  
Similarly, their observations would have been sensitive to a 10 M$_{Jup}$ 
planet with a projected separation of 8.5 AU or beyond for half their sample, 
and greater than 36 AU for the full sample.  These results (obtained by 
adopting published ages for the target stars, and using the appropriate 
planet models of \citet{cond}) point to a rarity 
of giant planets at large separations from their parent stars.  In this paper 
we enlarge the sample, and consider the implications of 
our null result with 
respect to the full orbital parameters of potential planets.  We aim to set 
quantitative limits on the distribution of planets in semi-major axis space, 
and statistically rule out models of planet populations.

\section{Observations}

We begin with contrast plots (sensitivity to faint companions as a function 
of angular separation from the target star) from two surveys for extrasolar 
planets, 
using large telescopes and adaptive optics.  \citet{elena} carried out a 
survey of 28 young, nearby, late-type stars with the NACO adaptive optics 
system at the 8.2 meter Very Large Telescope (VLT).  These observations have 
exposure times of order 30 minutes, with stars being observed in the H 
or Ks bands.  Subsequent to these observations, a survey of 54 young, nearby 
stars of a variety of spectral types (between A and M) was conducted 
between 2003 and 2005, with the results reported in \citet{sdifinal}.  This 
second survey used the Simultaneous Differential Imager (SDI) 
at the 6.5 meter MMT and the 8 meter VLT, an adaptive optics observational 
mode that allows higher contrasts by 
imaging simultaneously in narrow wavelength regions surrounding the 
1.6 $\mu$m methane feature seen in cool brown dwarfs and expected in 
extrasolar planets \citep{lenzen,abdor}.  This 
allows the light from a hypothetical 
companion planet to be more easily distinguishable from the speckle noise 
floor (uncorrected starlight), as the two will have very different spectral 
signatures in this region.  This translates to higher sensitivity at smaller 
separations than the observations of \citet{elena}, which were conducted 
before the VLT SDI device was commissioned (see Fig. 14 of 
\citet{sdifinal} for a more detailed comparison of the two surveys).  
For most of these SDI targets, the star was 
observed for a total of 40 minutes of integration time, which includes a 
33 degree roll in the telescope's rotation angle, in order to separate 
super speckles--which are created within the instrument, and so will not 
rotate--from a physical companion, which will rotate on the 
sky \citep{scr1845}.

For both sets of target stars, contrast curves have been produced which 
give the 5$\sigma$\footnote{We note that for the SDI observations this 
threshold corresponds to independent 5$\sigma$ measurements in both the 
0$^\circ$ and 33$^\circ$ images, see \citet{sdifinal} for details.} noise in 
the final images as a function of radius from the 
target stars, and thus an upper limit on the flux of an unseen planet in the 
given filter of the observations.  As no planets were detected in either 
survey at the 5$\sigma$ level, we use these contrast curves to set upper 
limits on the population of extrasolar planets around young, nearby stars.

\subsection{Target Stars}

\begin{figure}
\epsscale{1}
\plotone{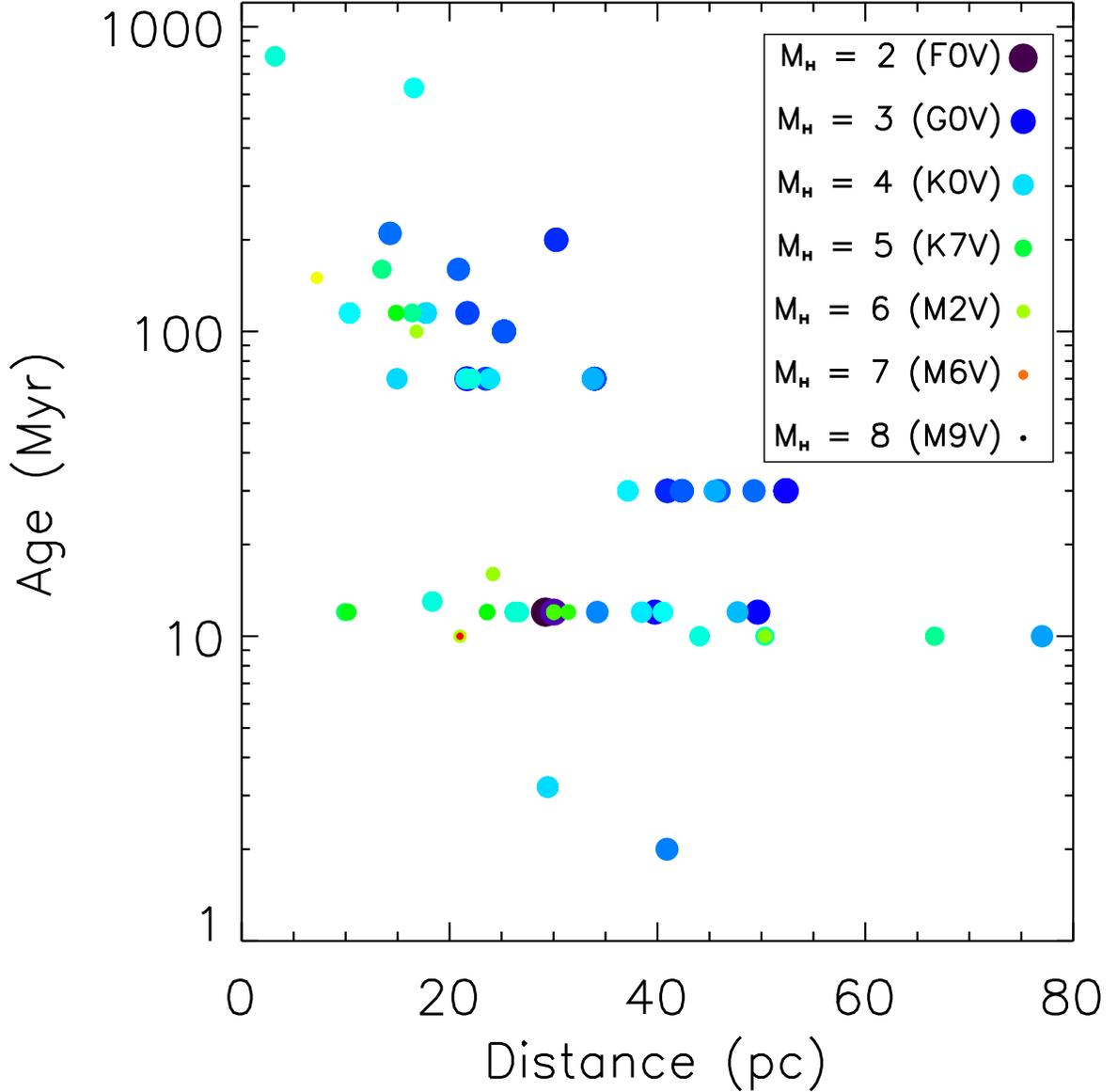}
\caption{The 60 target stars from our two surveys (though five stars are too 
old to appear on this plot).  These stars are some of the youngest, 
nearest stars known, spanning a range of spectral type.  The size of the 
plotting symbol and the color is proportional to the absolute H magnitude 
of the star: a bigger, bluer 
symbol corresponds to a brighter and hotter star.  The 
legend gives approximate spectral type conversions for main 
sequence stars, but we note that these stars have been plotted by their 2MASS 
H-band fluxes, and as a result their actual spectral type can vary from that 
shown in the legend.  See Table~\ref{table1} for more complete properties of 
these stars.  The median target star is a 30 Myr K2 star at 25 pc.
\label{targetsfig}}
\end{figure}

We construct a target list using 22 stars from the \citet{elena} survey, and 
48 stars from the survey of \citet{sdifinal}, for a total of 60 targets (10 
stars were observed by both surveys).  This 
first cut was made by considering stars from the two surveys that had contrast 
curves, and stars whose age could be determined by at least one of: group 
membership, lithium abundance, and the activity indicator R'$_{HK}$  (in three 
cases, ages from the literature were used, though these are stars that are 
generally older than our sample as a whole, and so uncertainties in the 
assumed ages will not adversely affect our results).  Ages are 
determined by taking the age of the moving group to which the target star 
belongs; if the star does not belong to a group, the lithium or R'$_{HK}$ age 
is used, or the two are averaged if both are available.  Lithium ages are 
found by comparing to lithium abundances of members of clusters of known ages, 
and similarly for R'$_{HK}$ \citep{mamajek}.  We give the full 
target list in Table~\ref{table1}, and details on the age determination in 
Table~\ref{table2}.  We also plot our target stars in Fig.~\ref{targetsfig}.  
Overall, our median survey object is a 30 Myr K2 star at 25 pc.

\section{Monte Carlo Simulations}

In order to place constraints on the properties of planets 
from our null results, we run a series of Monte Carlo simulations 
of an ensemble of extrasolar planets around each target star.  
Each simulated planet is given full orbital parameters, an 
instantaneous orbital phase, and a mass, then the planet's 
magnitude in the observational band is determined from these properties (using 
the target star's age and distance, and 
theoretical mass-luminosity relations) as is its projected separation from the 
star.  Finally, 
this magnitude is compared to the measured contrast curve to see if such a 
planet could be 
detected.  Determining which simulated planets were detected, and which 
were not, allows us to interpret the null result in terms of what models of 
extrasolar planet populations are excluded by our survey results.

\subsection{Completeness Plots}

As in \citet{sdifinal}, we use completeness plots to illustrate the 
sensitivity to planets as a function of planet mass and semi-major axis.  To 
do this, for each target star, we create a grid of semi-major axis and planet 
mass.  At each grid location we simulate $10^4$ planets, and then compute 
what fraction could be detected with the contrast curve for that star.

In general, most orbital parameters are given by well-known distributions.  
Inclination angle has a constant distribution in sin(i), while the 
longitude of the ascending node and the mean anomaly are given by uniform 
distributions between 0 and 2$\pi$.  Since contrast plots are given in terms 
of radius alone, it is not necessary to consider the argument of periastron 
in the simulations.

\begin{figure}
\epsscale{1}
\plotone{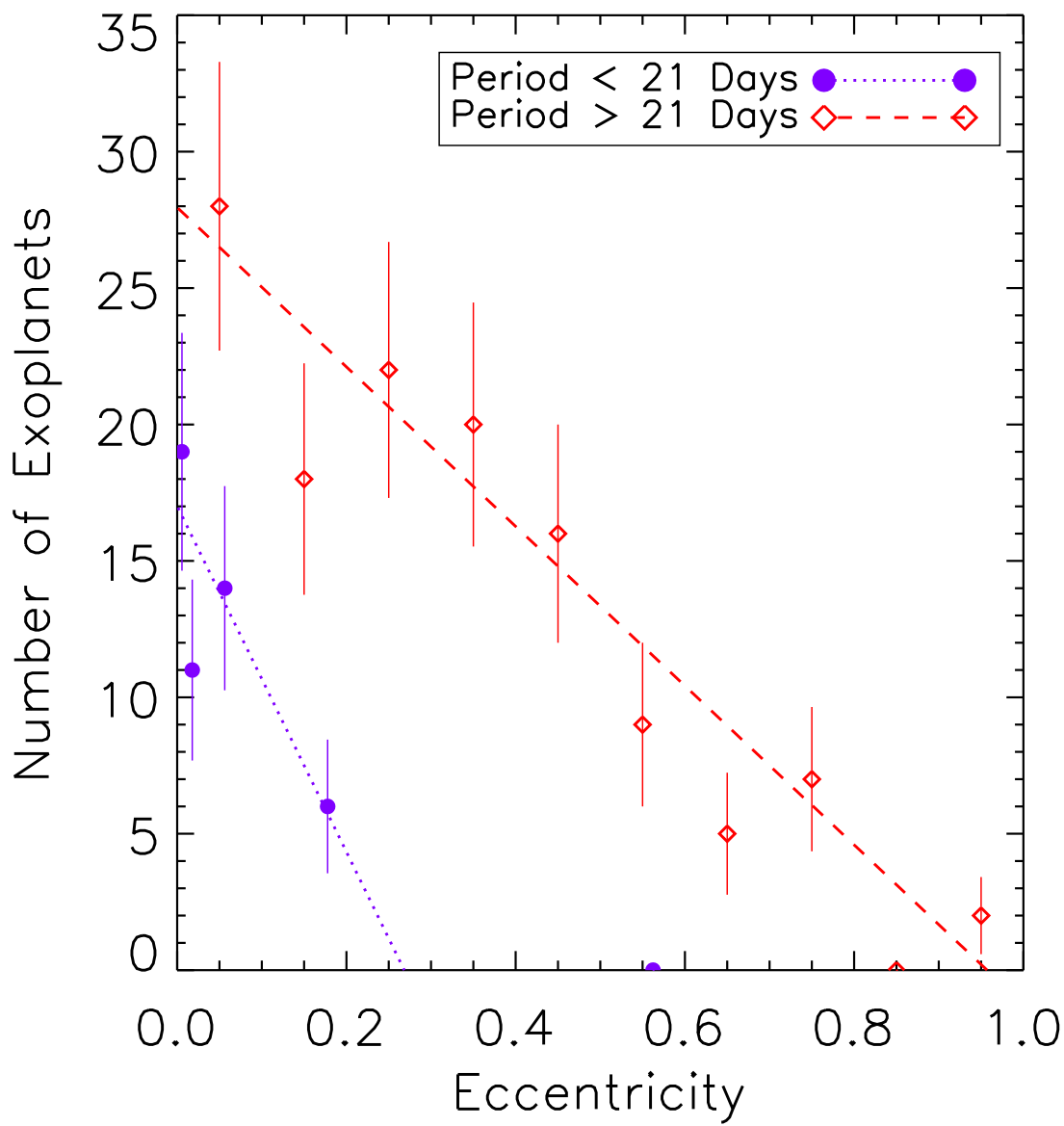}
\caption{The assumed distribution for the orbital eccentricities of 
extrasolar planets.  The datapoints represent the histograms for planets 
found to date with the radial velocity method  \citep{rvref}, with error bars 
as 1-sigma Poisson noise based on the number of planets per bin.  Planets are 
divided to separate ``Hot Jupiters,'' based on a period cut at 21 days; long 
period planets are divided into linear bins, short-period ones into 
logarithmic bins.  In both cases, a simple linear fit is a good representation 
of the data.  \label{eccfig}}
\end{figure}

To simulate the eccentricities of the planet orbits, we 
examine the orbital parameters of known extrasolar planets from radial 
velocity surveys.  We consider the orbits of planets given by \cite{rvref}, 
and show their distribution of eccentricities in Fig.~\ref{eccfig}.  By 
dividing the sample into two populations, based on a cut at an orbital period 
of 21 days, we can separate out the population of Hot Jupiters, which we 
expect to have experienced orbital circularization as a result of their 
proximity to their host stars.  For both sets of populations, we fit a simple 
straight line to the distributions (the logarithmic bins for the Hot Jupiter 
population means this line translates to a quadratic fit).  We note that the 
Hot Jupiter fit is plagued by small number statistics, and so the fit is 
likely to be less reliable than that for long period planets.  Even for our 
closest target 
stars, such an orbital period gives star-planet separations less than 0.1'', 
a regime where our contrast curves show we are not sensitive to planets.  As 
a result, the manner in which the orbits of Hot Jupiters are simulated has 
effectively no impact on our final results.

For each simulated planet, the on-sky separation is determined at the given 
orbital phase, and mass is converted into absolute H or Ks 
magnitude, following the mass-luminosity relations of both \citet{burrows} 
and \citet{cond}; both of these sets of models have shown success at 
predicting the properties of young brown dwarfs (e.g. \citet{stassun} and 
\citet{sinfoni}).  In the case of the models of \citet{burrows}, we use a 
Vega spectrum to convert the various model spectra into absolute H and Ks 
magnitudes.  We also note that the \citet{burrows} models only cover a 
range of planet 
masses greater than 1 $M_{Jup}$, and ages above 100 Myr.  Since the range of 
ages of our target stars extends down to 2 Myr, and we wish to consider 
planets down to masses of 0.5 $M_{Jup}$, we perform a simple extrapolation of 
the magnitudes to these lower ages and masses.  While this solution is clearly 
not ideal, and will not reflect the complicated physical changes in these 
objects as a function of mass and age, we feel that this method provides a 
good estimation of how the \citet{burrows} models apply to our survey.

At this point, we use the distance to the target star, as well as its 2MASS 
flux in either H or Ks, to find the delta-magnitude of each simulated 
planet.  With this, and the projected separation in the plane of the sky, we 
can compare each simulated planet to the 5$\sigma$ contrast curve, and 
determine which planets 
can be detected, and which cannot.  We also apply a minimum flux limit, based 
on the exposure time of the observation, as to what apparent magnitude for a 
planet is required for it to be detected, regardless of its distance from the 
parent star.  For the SDI observations, which make use of optimized (compared 
to basic H-band observing) methane filters, we add an additional 
factor of $\Delta$H$=$0.6 magnitudes (appropriate for a T6 spectral type, a 
conservative 
estimate for young planets; see \citet{sdifinal} for details on this factor).  
Also, for these SDI observations, we place 
an upper cut-off on masses where, for the age of the system, the planet 
reaches an effective temperature of 1400 K.  
Above this temperature, methane 
in the atmosphere of the 
planet is destroyed, and the methane feature disappears, so 
that the SDI subtraction now attenuates any planets, as well as stellar 
speckles.  While non-methane objects further out than 0.2'' are not totally 
removed in the image (e.g. Fig. 4 of \citet{abdorme}), for consistency we 
ignore this possibility when considering upper limits.

In many cases in our survey, a single target star was observed at several 
epochs, in some cases with different observational parameters (such as VLT 
NACO SDI and VLT NACO Ks broadband) or even different telescopes (MMT and 
VLT).  As a result, to be considered a null detection, a simulated planet must 
lie below the 5$\sigma$ detection threshold at each observational epoch, and 
this threshold must reflect the appropriate contrast curve for the given 
observation.  To account for this, for target stars with multiple 
observations, an ensemble of simulated planets is created for the earliest 
observational epoch, as described above, and compared to the contrast curve 
for that observation.  The simulated planets then retain all the same orbital 
parameters, except for orbital phase which is advanced forward by the 
elapsed time to the next observational epoch, and the simulated planets are 
now compared to the contrast curve from the later epoch (and so on for every 
available contrast curve).  A planet that lies above the contrast curve at any 
epoch is considered detectable.  Typically this elapsed time is about a year, 
and so is a minor effect for planets with long-period orbits; we nevertheless 
include this complexity for completeness.  The major benefit of this method is 
that for stars observed both with SDI \citep{sdifinal} and at H or Ks 
\citep{elena}, it is possible to leverage both the higher contrasts at smaller 
separations with SDI and the insensitivity to the methane feature of broadband 
imaging, which allows planets of higher masses to be accessed.  The epochs 
used when considering each observation are those given in Table 3 of 
\citet{elena} and Tables 2 and 3 of \citet{sdifinal}.

\begin{figure}
\epsscale{1}
\plotone{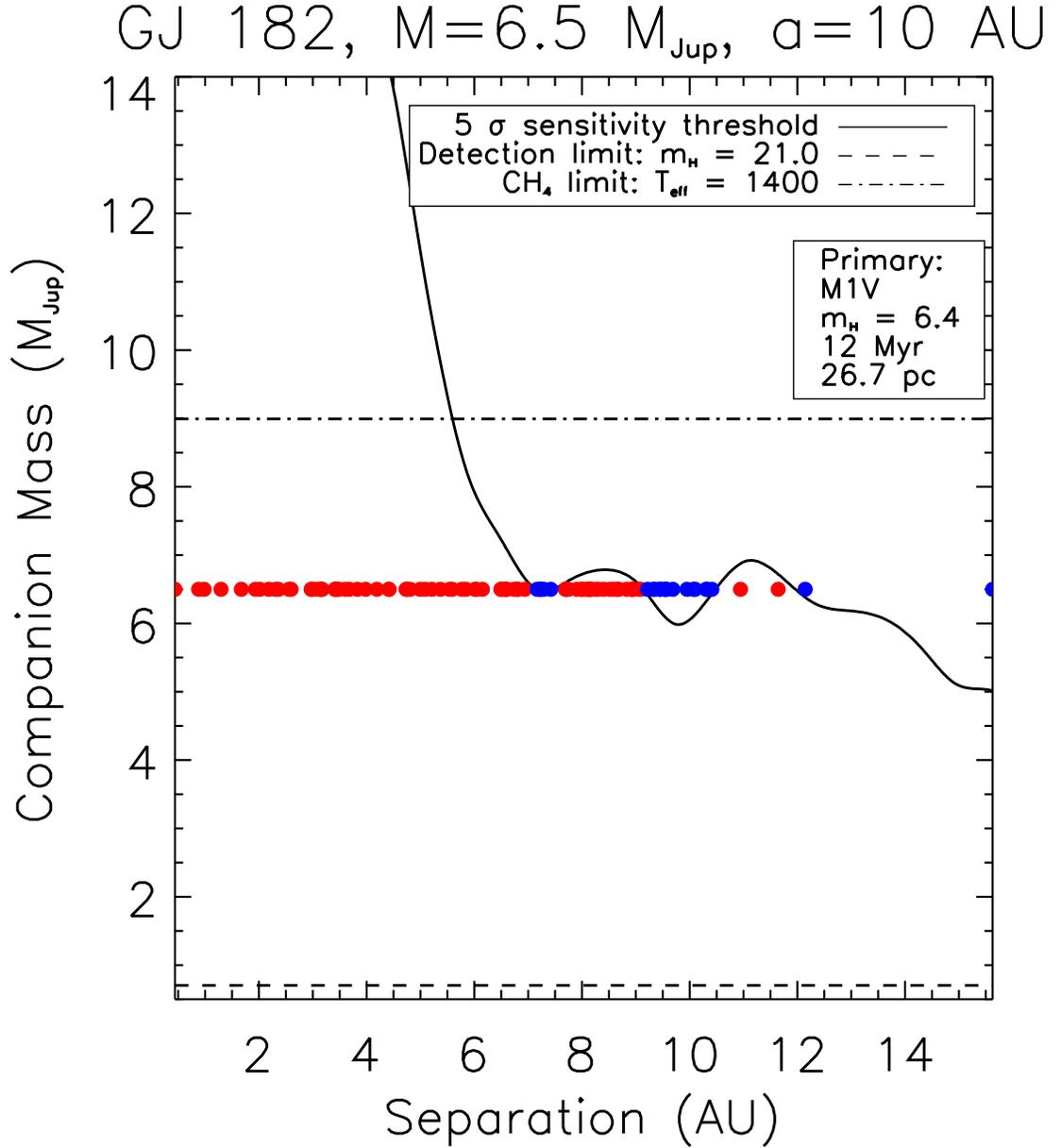}
\caption{The results of a single simulation of 10$^4$ planets around the 
SDI target star GJ 182 (for clarity, only 100 points are plotted here) 
\citep{sdifinal}.  Each planet has a mass of 
6.5 $M_{Jup}$, and a semi-major 
axis of 10 AU.  Due to various values of eccentricity, viewing angle, and 
orbital phase, the projected separation of each simulated planet departs from 
the semi-major axis, and the points smear across the horizontal direction, 
with projected separation running between 0 and 14 AU.  
Planets that are above the contrast curve are detected (blue dots), while 
those below are not (red dots).  In this case, 
20\% of these simulated planets were detected.  By 
running this simulation over multiple grid points of mass and semi-major axis, 
we produce a full completeness plot, such as Fig.~\ref{conpabfig}.
  \label{conpabdemofig}}
\end{figure}

We plot an example of this simulation at a single grid point in mass and 
semi-major axis in Fig.~\ref{conpabdemofig}, for the target star GJ 182, 
the 18th best target star in our survey, using the planet models of 
\citet{burrows}.  10$^4$ simulated planets (only 100 are plotted in this 
figure, for clarity) are 
given a single value of mass (6.5 $M_{Jup}$) and semi-major axis (10 AU).  
Since each planet has unique orbital parameters (eccentricity, viewing angle, 
and orbital phase), the projected separation varies from planet to planet, so 
some are above the 5$\sigma$ detection threshold of the contrast curve (the 
blue dots in the figure), while 
others are not (the red dots).  For this particular target star and 
simulated planets of mass 6.5 M$_{Jup}$ and semi-major axis 10 AU, 20\% of 
these planets can be detected.

\begin{figure}
\epsscale{.90}
\plotone{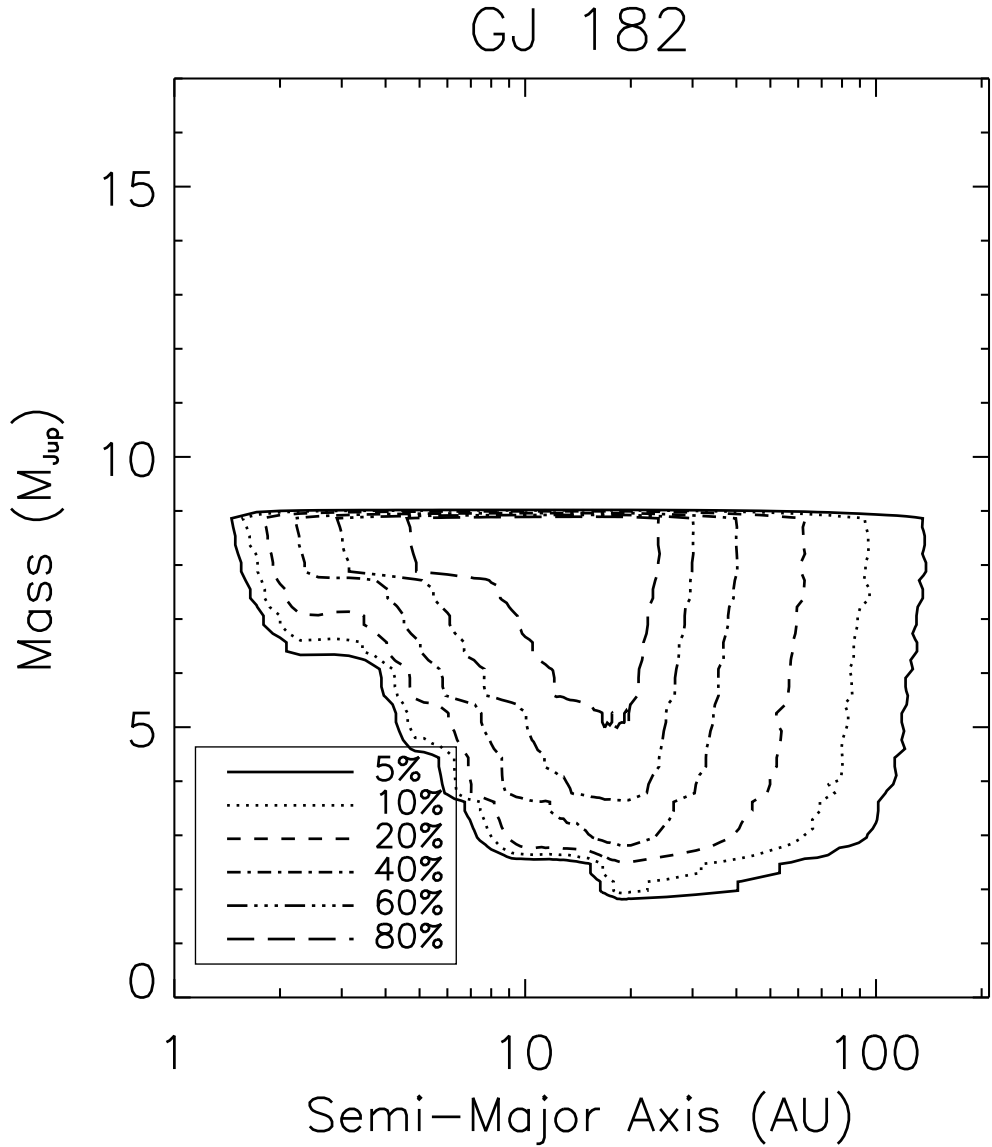}
\caption{A full completeness plot for the target star GJ 182.  As a function 
of planet mass and semi-major axis (with grid points between 0.5 and 17 
M$_{Jup}$ for mass, and semi-major axis between 1 and 4000 AU, though only 
the inner 210 AU are plotted here), the contours give the probability of 
detecting a planet with those parameters given the available contrast curve.  
At each grid point, 10$^4$ planets are simulated, as shown in 
Fig.~\ref{conpabdemofig}, and the fraction that can be detected is returned.  
The left edge is strongly influenced by the shape of the contrast curve, while 
the right edge depends mainly on the projected field of view of the 
observation.  The hard upper limit at 9 M$_{Jup}$ is set by the methane 
cut-off imposed by the SDI method, when the simulated planets exceed 1400 K 
and cease to have a strong methane signature in the spectrum.  The fact that 
the contours do not precisely line up at this limit is simply a result of the 
interpolation used to plot the contours.  Completeness plots for all 60 survey 
stars are available online at 
http://exoplanet.as.arizona.edu/$\sim$lclose/exoplanet.html
  \label{conpabfig}}
\end{figure}

To produce a complete contour plot, we consider a full grid of mass 
(100 points, 
between 0.5 and 17 $M_{Jup}$) and semi-major axis (200 points, 
between 1 and 4000 AU), running a simulation as in Fig.~\ref{conpabdemofig} at 
each of the 20,000 grid points.  We then plot contours showing what fraction 
of planets we can detect that have a given mass and semi-major axis, 
in Fig.~\ref{conpabfig}, again for the target star GJ 182.  The hard upper 
limit is set by the methane cut-off, where the planet mass becomes high 
enough (for the age of the given target star) for the effective temperature 
to exceed 1400 K, at which 
point the methane feature is much less prominent in the planet's spectrum.  
Although there exists a Ks dataset for the star GJ 182, and additional 
observational epochs with SDI, for clarity we only use a single SDI contrast 
curve to produce this figure; the full dataset is used for subsequent 
analysis.  If GJ 182 had a planet with mass and semi-major axis such that it 
would fall within the innermost contour of Fig.~\ref{conpabfig}, we would have 
an 80\% chance of detecting it.  Obviously, 
these plots make no statements about whether these stars have planets of the 
given parameters, but instead simply express our chances of detecting such a 
planet if it did exist.

\subsection{Detection Probabilities Given an Assumed Distribution of Mass and Semi-major Axis of Extrasolar Planets}

\begin{figure}
\epsscale{1}
\plotone{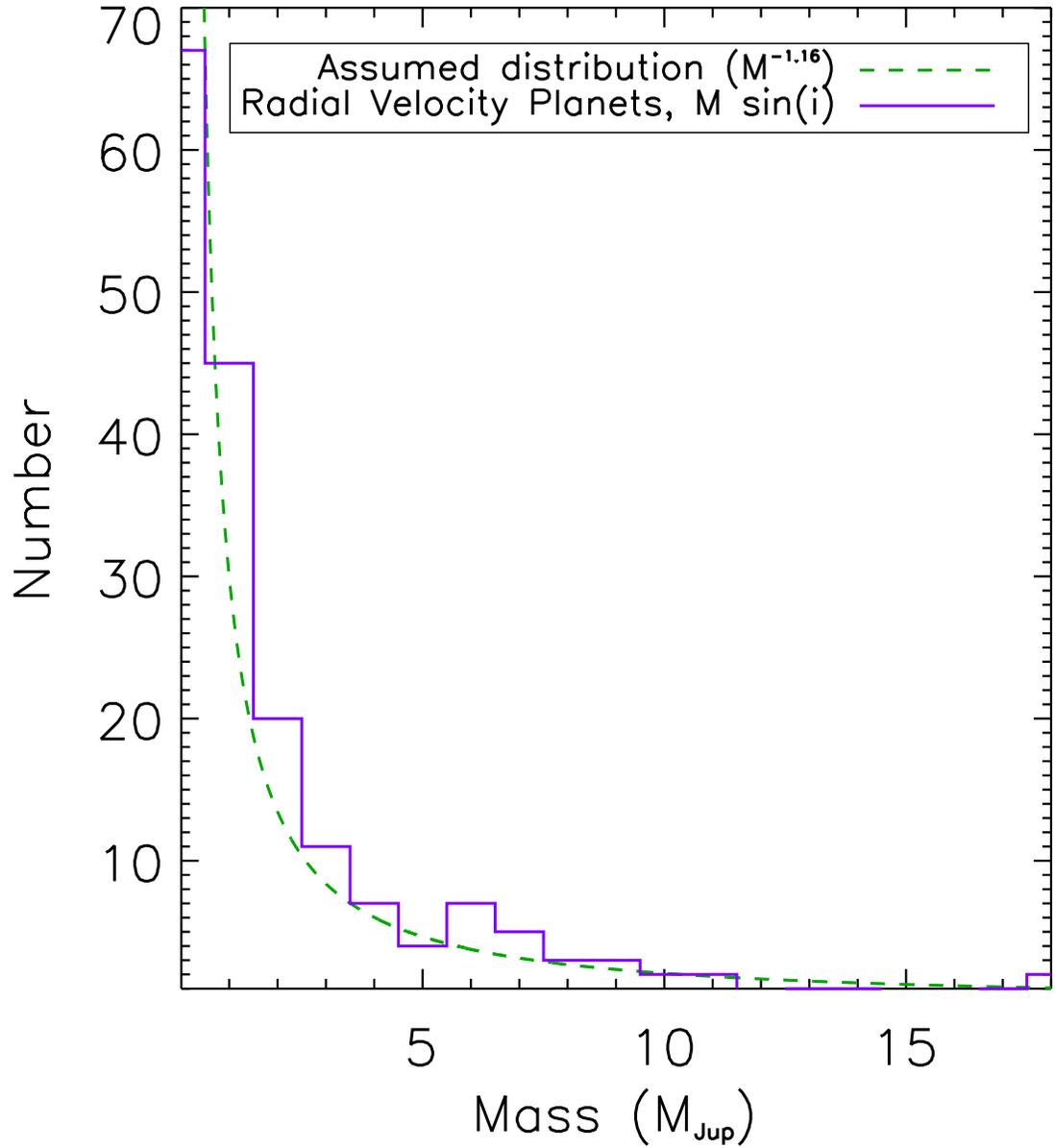}
\caption{The assumed mass distribution of extrasolar planets, plotted against 
the histogram of known planets detected by the radial velocity method.  
Throughout this paper we 
adopt a power law of the form $\frac{dN}{dM} \propto M^{-1.16}$, as suggested 
by \citet{rvref}, which does a reasonable job fitting the data.
  \label{massfig}}
\end{figure}

\begin{figure}
\epsscale{1}
\plotone{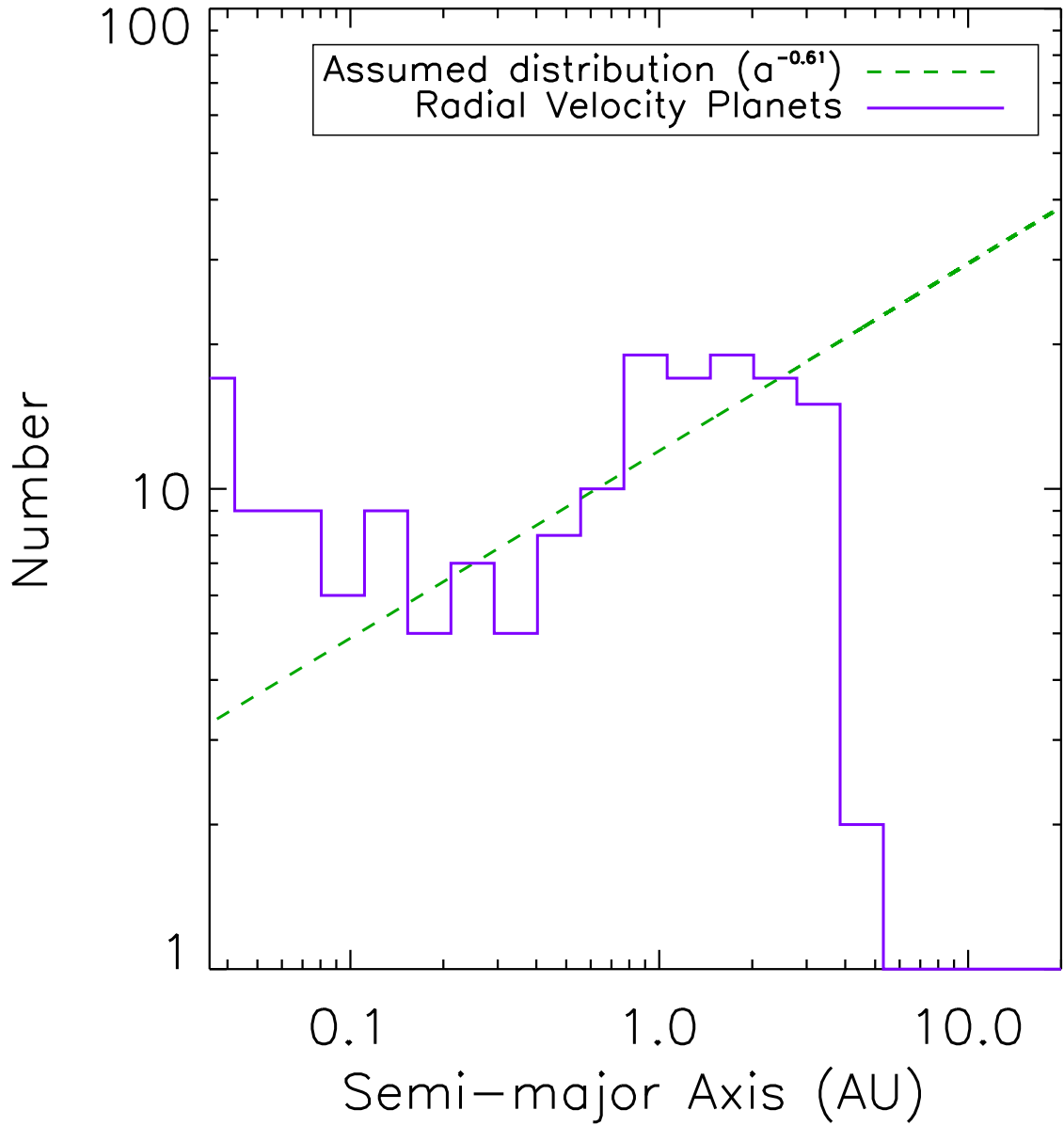}
\caption{The distributions that we consider for semi-major axis of 
extrasolar planets, again with the histogram of known radial velocity 
planets.  We adopt the observed distribution of \citet{cumming}, with 
$\frac{dN}{da} \propto a^{-0.61}$, which is suggestive of the existence of 
wider planets, given that 
radial velocity surveys should be especially sensitive to hot Jupiters 
(producing an over-abundance at small separations) and less sensitive to 
long-period orbits (resulting in a decline in detected planets at larger 
separations).
  \label{smafig}}
\end{figure}

With the large number of currently-known extrasolar planets, it is possible 
to assume simple power-law representations of the distributions of mass and 
semi-major axis of giant planets, which allows for a more quantitative 
interpretation of our null result.  \citet{rvref} suggest a power law of 
the form $\frac{dN}{dM} \propto M^{-1.16}$ for mass, while \citet{cumming} 
use the power law $\frac{dN}{da} \propto a^{-0.61}$ for semi-major 
axis, in order to describe the distributions of known extrasolar planets.  We 
make histograms for 
mass (Fig.~\ref{massfig}) and semi-major axis (Fig.~\ref{smafig}) from the 
parameters of all currently-known extrasolar planets (parameters taken from 
the Catalog of Nearby Exoplanets, http://exoplanets.org, in May 2007).  In 
both cases, the power laws do a reasonable job of fitting the data, above 
1.6 $M_{Jup}$ and within 3.5 AU.  For smaller planets, or longer periods, we 
would expect the observational biases of the radial velocity method to make 
the sample incomplete, thus accounting for the drop-off of planets from what 
would be predicted by the power law.  We echo the caution of \citet{rvref} 
that these planets are drawn from many inhomogeneous samples, but we believe 
with the relatively large numbers the derived distributions are not far off 
from the actual distributions.

In general, then, if one assumes that these power laws are universal to all 
stars, and that the semi-major axis power law continues to larger separations 
with the same index, the only outstanding question is to what outer limit 
(or ``cut-off'') this distribution continues before it is truncated.  This 
cut-off is a term that can be uniquely well constrained by the null results 
from our survey.  We return to this 
issue, after considering the results from our survey, in Section~\ref{smasec}.

\begin{figure}
\epsscale{.8}
\plotone{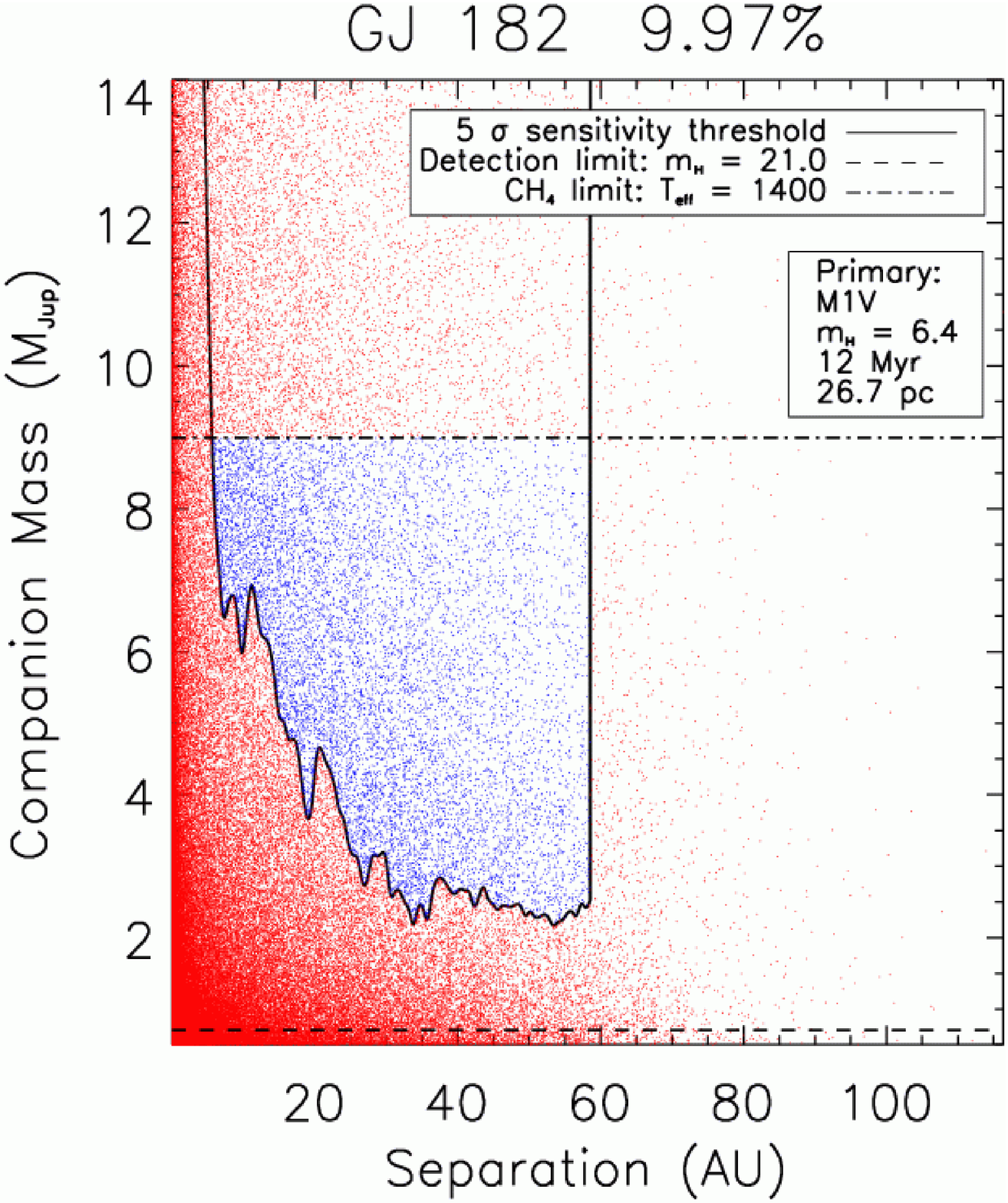}
\caption{10$^5$ simulated planets around the SDI target star GJ 182, following 
the distributions for mass ($\frac{dN}{dM} \propto M^{-1.16}$) of 
\citet{rvref} and semi-major axis ($\frac{dN}{da} \propto a^{-0.61}$) of 
\citet{cumming}, with mass 
running from 0.5 to 13 $M_{Jup}$, and semi-major axis cut off at 70 AU (since 
there is a range of eccentricities, separation can exceed the semi-major axis 
cut-off).  Detected planets (blue dots) are those that lie above the contrast 
curve, 
above the minimum flux level, and below the methane cut-off.  In this case, 
10\% of the simulated planets could be detected with this observation.  Using 
the metric of completeness to planets with this mass and semi-major axis 
distribution, GJ 182 is the 18th best target star in our sample.
\label{msplotfig}}
\end{figure}

For the Monte Carlo simulations using these assumptions, in addition to the 
other orbital parameters, we obtain mass and 
semi-major axis through random variables that follow the given power law 
distributions, and again find what fraction of planets can be detected given 
the contrast curve for that particular target star.  An example of this 
simulation, again for GJ 182, is given in Fig.~\ref{msplotfig}, showing that 
with an assumed upper limit for semi-major axis of 70 AU, and a power law 
with index -0.61, and mass power law index of -1.16 between 0.5 and 13 
M$_{Jup}$, we would be able 
to detect 10\% of the simulated planets.  Again, for this figure, we simply 
show the results using the models of \citet{burrows}.

\section{Analysis}

Having developed the tools to produce completeness plots, as well as compute 
the fraction of detected planets for various assumed models of semi-major 
axis, we proceed to combine the results over all our target stars in order to 
place constraints on the populations of extrasolar planets from these two 
surveys.

\subsection{Planet Fraction}

A simplistic description of the number of planets expected to be 
detected is given by the expression

\begin{equation}
N(a,M) = \sum^{N_{obs}=60}_{i=1} f_p(a,M) P_{i}(a,M)
\label{planeteq}
\end{equation}

\noindent That is, the number of planets one expects to detect at a certain 
semi-major axis and mass is given by the product of the detection probability 
($P_i$) for a planet of that mass (M) and semi-major axis (\textit{a}), and 
the fraction of stars ($f_p$) that contain such a planet (or 
``planet fraction''), summed over all target stars.  In this treatment, 
we ignore two major effects: we assume that there is no change in the mass or 
separation distribution of planets, or their overall frequency, as a function 
of spectral type of the primary; we also do not consider any metallicity 
dependence on the planet fraction.  While these assumptions are clearly 
incorrect (e.g. \citet{johnson}, \citet{fv05}), it is a good starting point 
for considering what constraints can be placed on the population of 
extrasolar planets.  Also, we note that our sample includes 24 binaries, 
which may inhibit planet formation, though most of these binaries have 
separations greater than 200 AU.  
This leaves only ten binaries with separations in the range of likely planet 
orbits that might potentially 
contaminate our results.  For simplicity, we leave these binaries in our 
sample, and we will return to this issue in Section~\ref{smasec}.

Using the contrast curves from each of our 60 targets stars (as in 
Fig.~\ref{conpabfig}), we simply sum the fraction of detectable planets at 
each grid points for all of our stars.  This gives the predicted number of 
detectable 
planets at each combination of mass and semi-major axis, assuming each target 
star has one planet of that mass and semi-major axis ($f_p(a,M)$=1).

More instructively, if we assume a uniform value 
of the planet fraction for all target stars, we can solve for $f_p$.  Then by 
assuming a particular value for the predicted number of planets 
($\Sigma P_i$), our null result allows us to place an upper limit on the 
planet fraction at a corresponding confidence level, since our survey 
measured a value of N(a,M)=0.  In a Poisson distribution, the probability of 
obtaining a certain value is given by $P=e^{-\mu} \frac{\mu^\nu}{\nu!}$, which 
for the case of a null result, $\nu=0$, becomes $P=e^{-\mu}$,  so a 95\% 
confidence level requires an expectation value, $\mu$, of 3 planets.  We can 
thus rewrite Eq.~\ref{planeteq}, using $\Sigma P_i = 3$, as 
\begin{equation}
f_p(a,M) \le \frac{3}{\sum^{N_{obs}}_{i=0} P_{i}(a,M)}
\label{planeteq2}
\end{equation}
\noindent Put another way, if we expected, from our 5$\sigma$ contrast 
curves, to detect 
12 planets ($\Sigma$P=12, for $f_p = 1$), in order to have actually detected 
0 planets from our entire survey (N=0), the planet fraction must be less 
than $\frac{3}{12}$ = 25\% ($f_p<$0.25), at the 95\% confidence level.  Doing 
this at each point in the grid of our completeness plots allows for an upper 
limit on the planet fraction as a function of mass and semi-major axis.

\begin{figure}
\epsscale{1}
\plotone{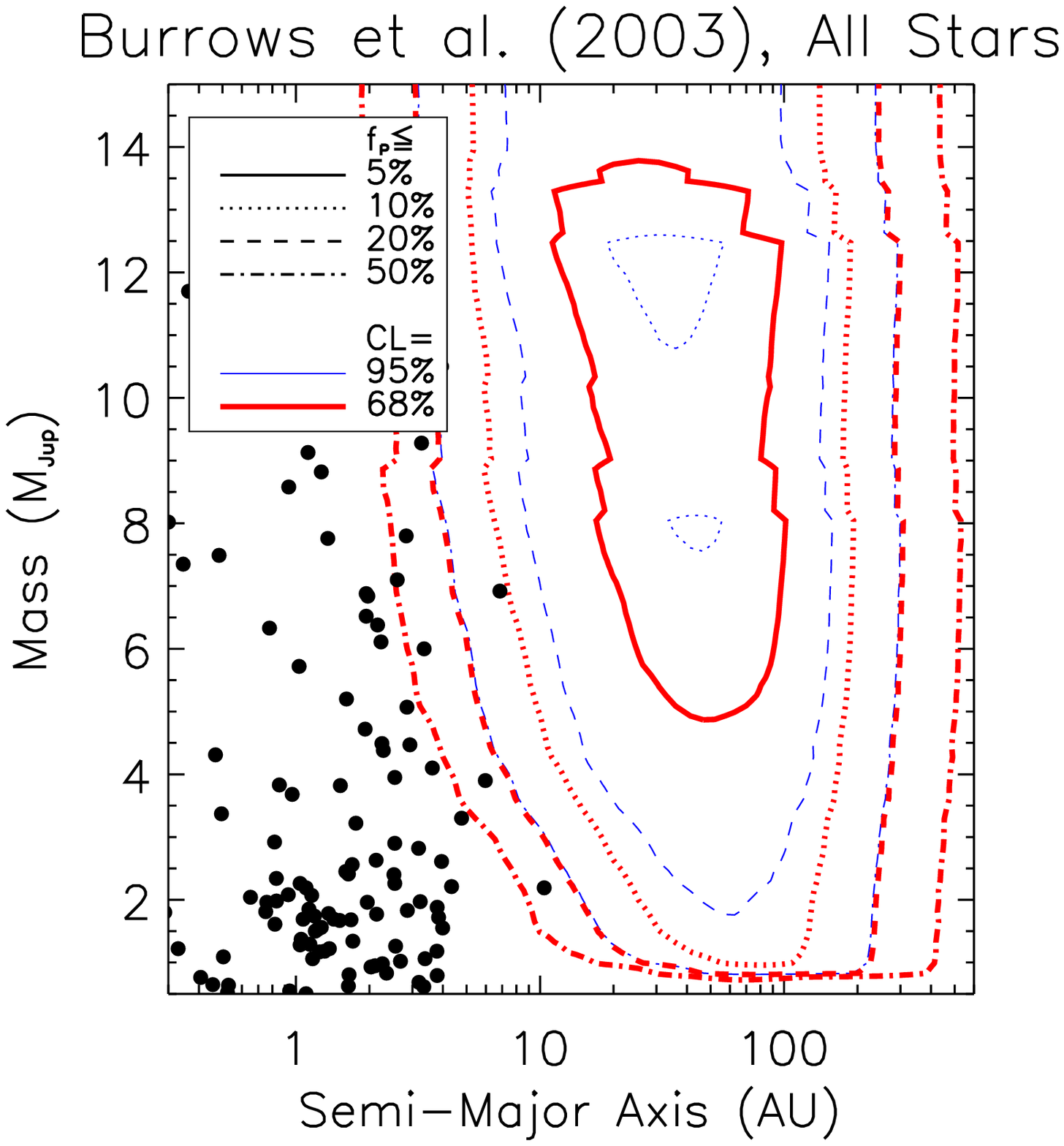}
\caption{The upper limit on the fraction of stars with planets ($f_p$), as a 
function of mass and semi-major axis (see Eq.~\ref{planeteq2}), using the 
planet models of \citet{burrows}, with the 95\% confidence level plotted as 
thin blue lines.  
We also plot in thicker red lines the 68\% confidence level contours.  
Given the results of our survey, we would expect, for 
example, less than 20\% (as indicated by the thin dashed blue line) of stars 
to have a planet of mass greater than 4 $M_{Jup}$ in an 
orbit 20 $<$ a $<$ 100 AU, and less than 50\% of stars (the dot-dashed thin 
blue line) to have planets more 
massive than 4 $M_{Jup}$ with semi-major axes between 8 and 250 AU, at the 
95\% confidence level.  Also 
plotted in the solid circles are known extrasolar planets.  There is still 
a gap between planets probed by direct imaging surveys (such as the ones 
described in this work), and those using the radial velocity method.
  \label{contourfig}}
\end{figure}

We plot the contours of this upper limit in Fig.~\ref{contourfig}, using the 
planet models of \citet{burrows}.  
A general result from 
these data is that, again at the 95\% confdience level, we would expect fewer 
than 20\% of stars to have planets of mass greater than 4$M_{Jup}$ with 
semi-major axis between 20 and 100 AU.  There appears to be no ``oasis'' of 
giant planets (more massive than Jupiter) in long-period orbits: at the 85\% 
confidence level, this upper limit on the fraction of stars with 
giant planets drops to less than 10\%.

\begin{figure}
\epsscale{1}
\plotone{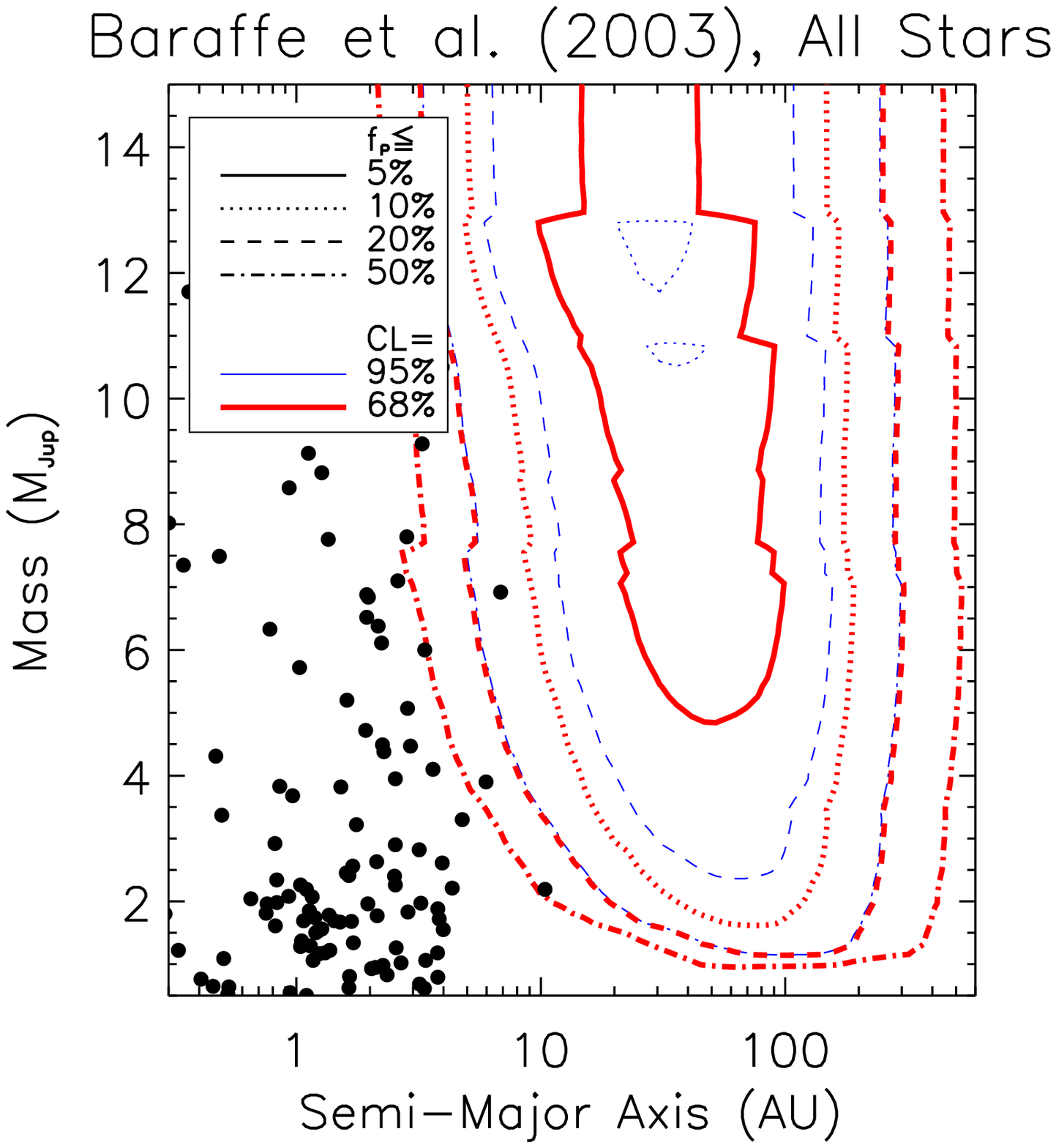}
\caption{The same as Fig.~\ref{contourfig}, but instead using the models 
of \citet{cond} to convert between planet mass and NIR magnitudes.  The COND 
models generally predict brighter planets for higher masses, but fainter 
planets at lower masses, compared to the \citet{burrows} models.  
Nevertheless, the two sets of models predict similar overall results.
\label{contourfiglyon}}
\end{figure}

We present the same plot, this time 
using the COND models of \citet{cond}, in Fig.~\ref{contourfiglyon}.  As the 
two sets of models predict quite similar planet NIR magnitudes, the plots 
are virtually the same.  The main difference between these models is that, 
given the age distribution of our target stars, 
higher mass planets appear slightly brighter in the \citet{cond} models, 
with the trend reversing and lower mass planets becoming fainter, as compared 
to the models of \citet{burrows}.
\citet{marley} have recently produced a third 
set of models, which globally predict lower luminosities for giant 
planets.  Since synthetic spectra for these models are not currently 
available, we do not examine the consequences of these models here, though 
we discuss possible effects in Section~\ref{discussion}.  But we note that 
while at 30 Myr and at 4 M$_{Jup}$ there is only a $\sim$3X decrease in the 
luminosity predicted by \citet{marley} compared to \citet{burrows}, 
the temperature of these objects is lower, therefore increasing the number 
of planets with methane that can be detected using SDI.  As a result, even 
with the future use of the \citet{marley} models, our results will not 
change dramatically, with respect to the total number of planets to which 
we are sensitive.

\subsection{Host Star Spectral Type Effects}

From the perspective of direct imaging searches for extrasolar planets, 
M-stars are especially appealing: their lower intrinsic luminosity means a 
given achievable contrast ratio allows fainter companions to be detected, 
and so makes the detection of planet-mass companions seem more likely.  
Nevertheless, there appears to be mounting evidence that even if the fraction 
of stars with planets does not decline when moving to later spectral types 
(and the work of \citet{johnson} suggests this fraction does decrease for 
M stars), 
the mean planet mass is likely to decrease (e.g. \citet{neptune_candc},
 \citet{neptune_swiss}).  While it seems natural that the initial mass of the 
circumstantial disk (and so the mass of formed planets) should scale with the 
mass of the parent star, such a relation is not easily quantified for planets 
at all orbital separations.  Additionally, it is problematic for us to model 
planet distributions for M star hosts on radial velocity planets, when these 
planets are almost entirely in systems with a host star of spectral type F, 
G, or K.

\begin{figure}
\epsscale{1}
\plotone{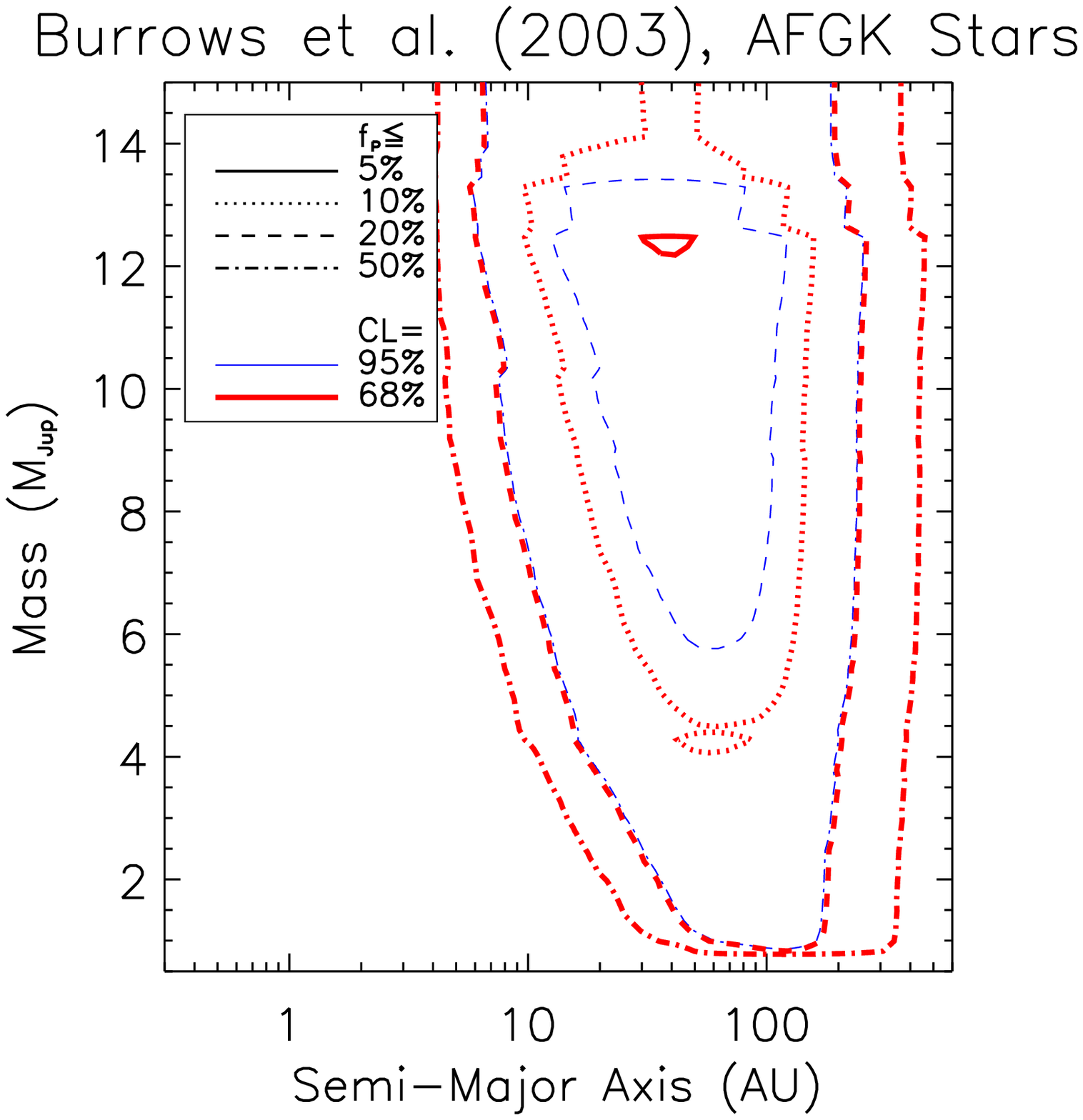}
\caption{The 95\% and 68\% confidence upper limit on planet fraction, 
limited only to stars of 
spectral type A through K, using the \citet{burrows} models.  Since with 
earlier spectral types the parent star is intrinsically brighter, it becomes 
more difficult to access planets of smaller masses or smaller separations.  
For AFGK stars we can only say, at the 95\% confidence level, that 
less than 20\% of stars have $M>7 M_{Jup}$ 
planets at 30-70 AU, or a limit of 50\% for planets with masses above 
6 $M_{Jup}$ at 10-200 AU.
\label{contourfigfgk}}
\end{figure}

\begin{figure}
\epsscale{1}
\plotone{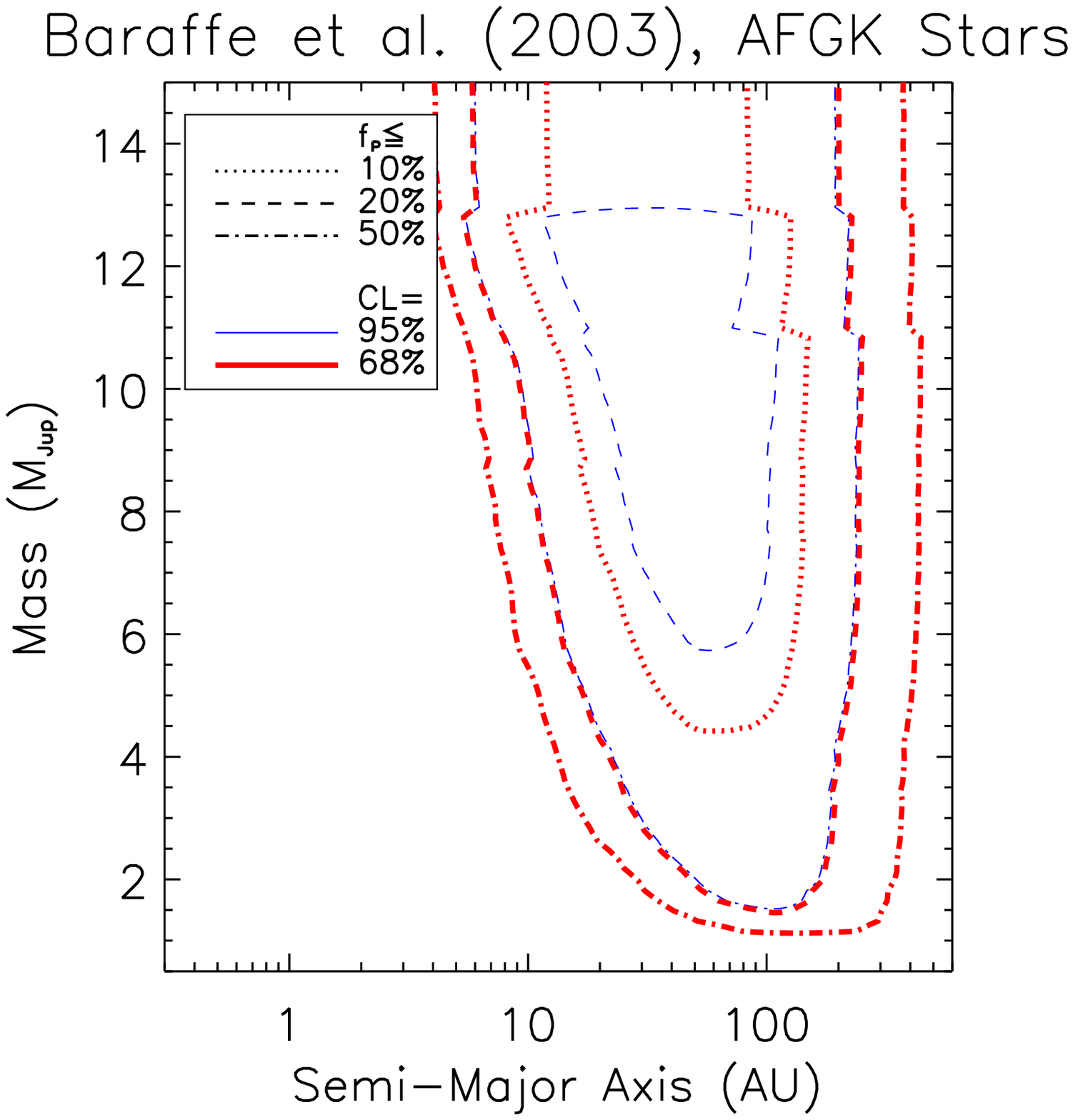}
\caption{The same as Fig.~\ref{contourfigfgk}, but with the \citet{cond} 
models used to find planet masses.
\label{contourfigfgklyon}}
\end{figure}

In order to investigate this effect, we divide our stars by spectral type, 
then recompute what limits we can set on the planet fraction.  In 
Fig.~\ref{contourfigfgk} and~\ref{contourfigfgklyon} we plot the upper limit 
on the planet fraction for 
only the solar-like stars (K or earlier) in our survey (45 of our 60 
target stars, this includes the one A star in our survey, HD 172555 A).  As 
we would expect, the statistics in the inner 
contour remain largely the same, but the contours move upward and to the 
right, as less massive and closer-in planets become harder to detect against 
the glare of earlier-type stars.

\begin{figure}
\epsscale{.85}
\plotone{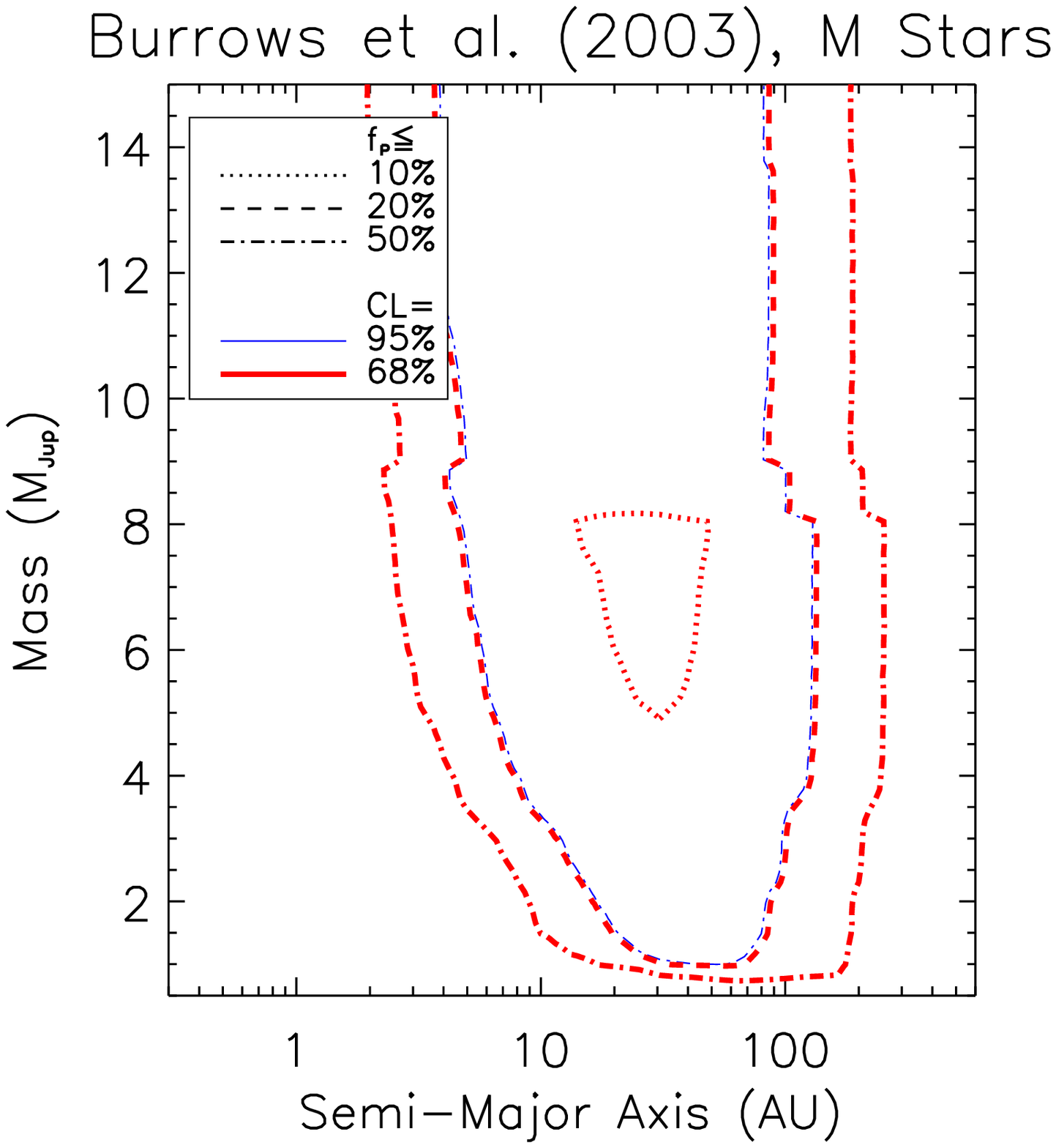}
\caption{Now using only our 15 M stars, we again plot 
the 95\% and 68\% confidence level upper limit on planet 
fraction, using the \citet{burrows} models.  While the plot follows the 
shape of Fig.~\ref{contourfig}, the removal of about three-quarters of the 
target 
stars reduces the upper limit that can be set on the planet fraction.  Hence 
less than 50\% of M stars should have planets with $M>4 M_{Jup}$ from 10 to 
80 AU, at 95\% confidence.  The analysis of microlensing results by 
\citet{microlens2} sets upper limits on the planet fraction of M dwarfs 
in the galactic bulge of 45\% for 3 M$_{Jup}$ planets between 1 and 7 AU, and 
33\% for 1 M$_{Jup}$ planets between 1.5 and 4 AU.  While even our 50\% 
contour (at the 68\% confidence level) does not probe the area of parameter 
space considered by \citet{microlens2}, which places upper limits on 1 
M$_{Jup}$ planets between 1.5 and 4 AU around M dwarfs of $\leq$33\%, and 
$\leq$45\% for 3 M$_{Jup}$ planets between 1 and 7 AU, 
the microlensing upper limits are unsurprising given our limits at somewhat 
larger separations for planets of the same mass.  Though we note that the 
composition (especially in terms of stellar metallicity) is 
likely to differ greatly between the two samples.  Also, we again draw 
attention to the fact that \citet{johnson} has shown that for M stars, giant 
planets at small radii are less common than around more massive stars.
\label{contourfigmstar}}
\end{figure}

\begin{figure}
\epsscale{1}
\plotone{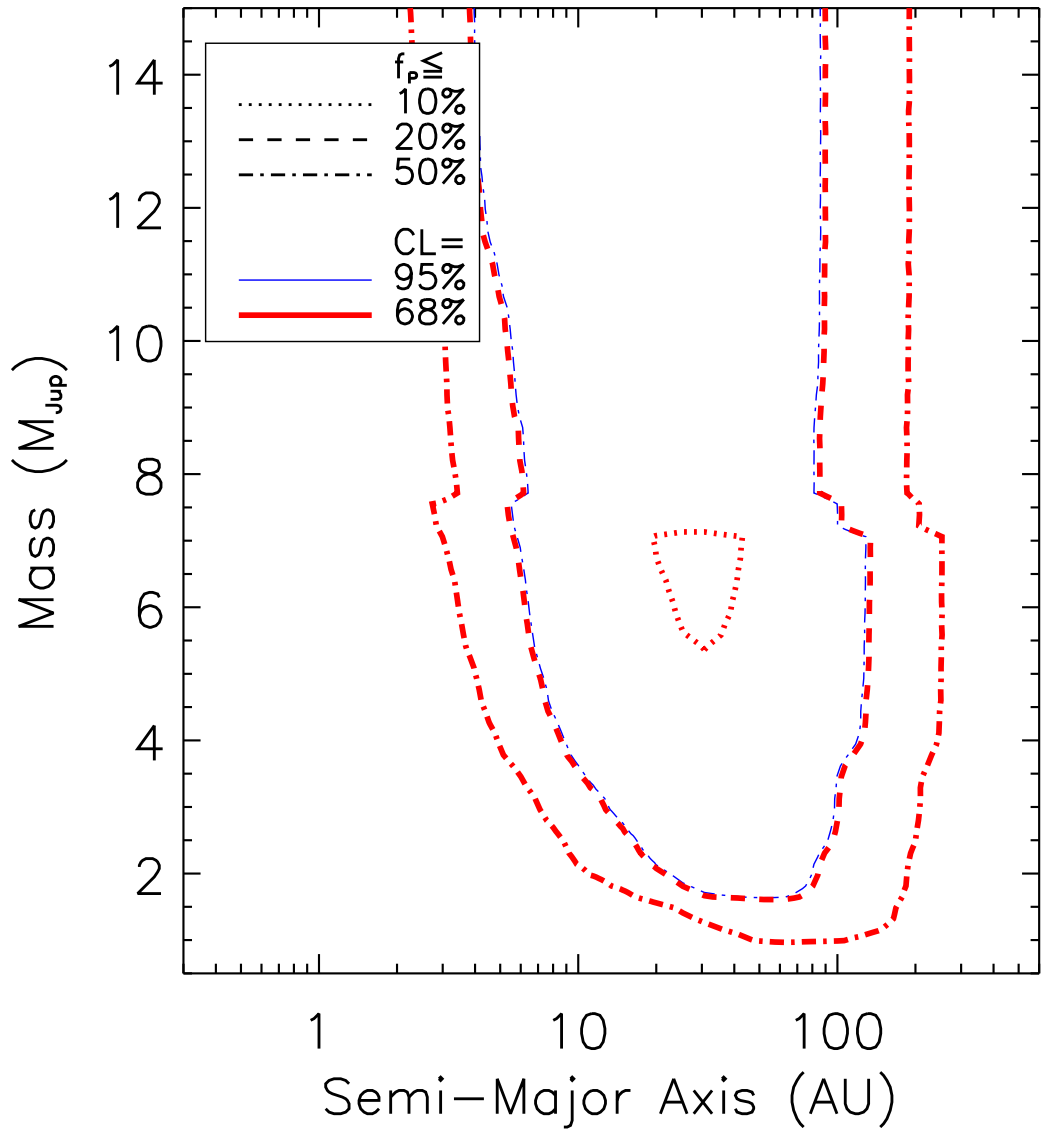}
\caption{As with Fig.~\ref{contourfigmstar}, only now with the \citet{cond} 
models used to find planet masses.
\label{contourfigmstarlyon}}
\end{figure}

We also consider the fifteen M stars in our sample, in 
Fig.~\ref{contourfigmstar} and~\ref{contourfigmstarlyon}.  The effect of 
the smaller number of stars is apparent, though the shape of the contours 
is again roughly the same.  If, as is suggested by \citet{johnson}, giant 
planets are less common around low mass stars, or less massive stars harbor 
less massive planets, it becomes difficult to probe the population of M star 
planets with surveys such as these.



\subsection{Constraining the Semi-Major Axis Distribution}\label{smasec}

We now consider what constraints can be placed on planet populations if 
we assume a basic form to the distributions.  In particular, if we take the 
mass power-law from currently-known extrasolar planets, 
$\frac{dN}{dM} \propto M^{-1.16}$ \citep{rvref}, we can 
constrain what types of power laws for semi-major axis are allowed by our 
survey null result.  To accomplish this, we simulate planets using a grid 
of power law indices and upper cut-offs for semi-major axis for each of our 
target stars.  Then, the sum of the detection fractions over the entire 
survey gives the expected number of detected planets, assuming each star 
has one planet (for example, if for 10 stars, we had a 50\% chance of 
detecting a planet around each star, we'd expect to detect 5 planets after 
observing all 10 stars).  Since we've set the distribution of planets, we can 
determine the actual planet fraction: radial velocity surveys tell 
us this value is 5.5\% for planets more massive than 1.6 $M_{Jup}$ and 
with periods shorter than 4 years (closer-in than 2.5 AU) \citep{fv05}.  
We can then use 
the mass and semi-major axis power laws to find the planet fraction for 
planets down to 0.5 Jupiter masses and out to the given semi-major axis 
cut-off, while always preserving the value of 5.5\% for the planet fraction 
for planets $>$1.6 M$_{Jup}$ and $<$2.5 AU.  Then, by multiplying this planet 
fraction by the sum of 
detection probabilities, we find the expected number of planets we'd detect 
given each distribution.  At this point, we can again use the Poisson 
distribution to convert this to a confidence level (CL) for rejecting the 
model, given our null result: $CL=1-e^{-\mu}$, where $\mu$ is the expected 
number of planets for that model.

Since stellar multiplicity is likely to disrupt planet formation, we exclude 
all known stellar binaries from our target list with projected separations 
less than 200 AU.  Since our results deal mainly with the inner 100 AU around 
our target stars, binaries that are any closer would greatly influence 
the formation of planets at these radii, creating an entirely different 
population.  \citet{exobinaries} have shown that while the overall planet 
fraction (for radial velocity planets, as taken from the volume limited 
sample of \citet{fv05}) is similar between single stars and wide binaries, it 
decreases for stars in tight binary systems.  Our inner cut-off on binary 
separation is at a larger separation than that noted in \citet{exobinaries}, 
but we consider planets in much wider orbits than those detectable with the 
radial velocity method.  Additionally, it has been shown by \citet{alphacen1} 
and \citet{alphacen2} that terrestrial planets could form and survive in the 
$\alpha$ Cen AB system, 
despite the relatively tight (23 AU), high eccentricity (0.5) orbit. 
\citet{alphacen2} also found that for most cases, a planet is stable in a 
binary system if its orbital radius is less than $\sim$10-20\% of the binary 
separation.  Applying this additional condition 
to our sample, we remove 1 star from the \citet{elena} survey, and 9 from 
the \citet{sdifinal} sample, leaving 50 stars in our sample.  We give further 
details on the binaries in our sample in Table~\ref{table3}.

\begin{figure}
\epsscale{1}
\plotone{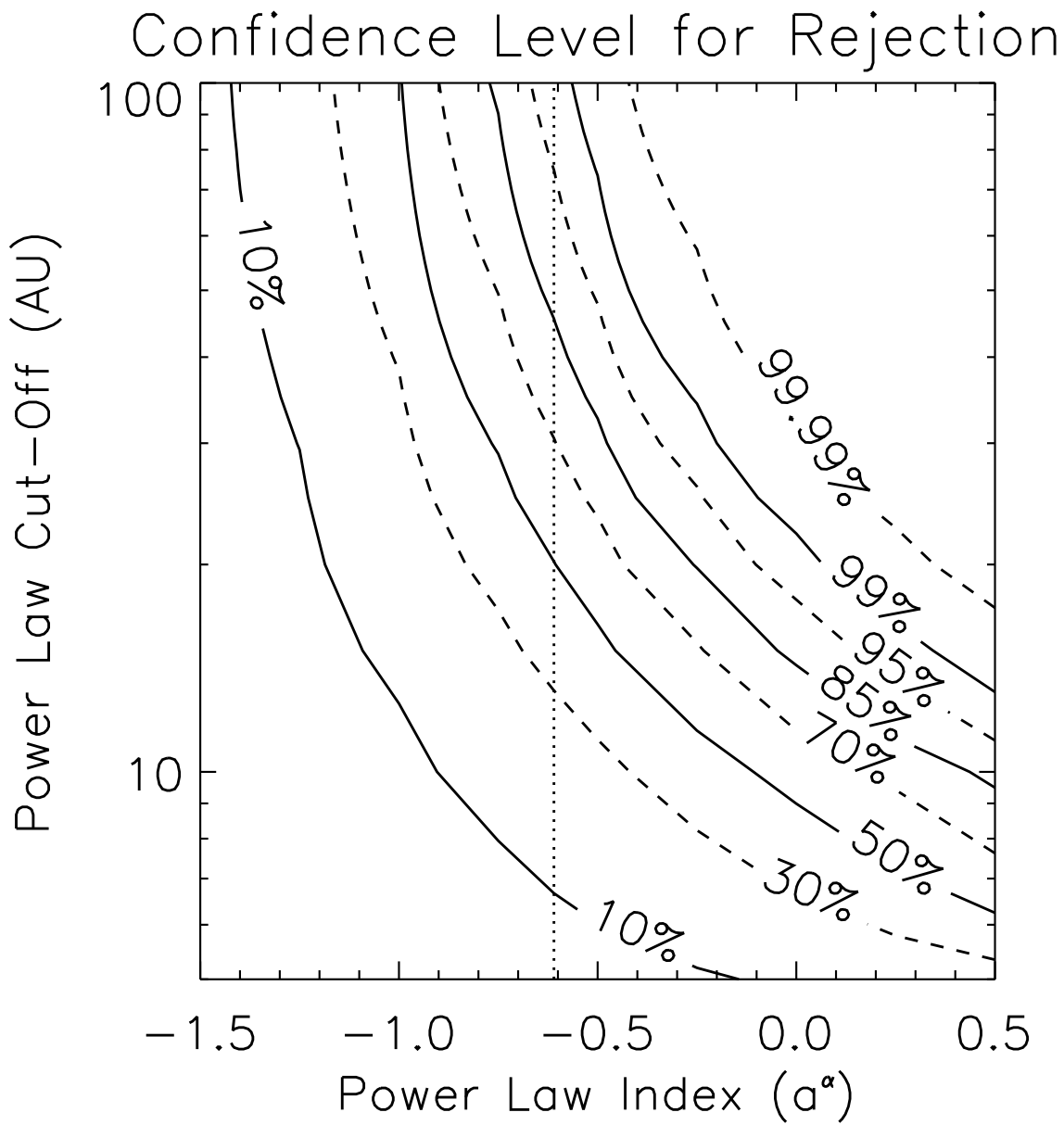}
\caption{The confidence level with which we can reject models of planet 
populations, assuming a power-law 
distribution for semi-major axes ($\frac{dN}{da} \propto a^{\alpha}$), as 
a function of the power law index and 
upper cut-off (N(\textit{a})=0 for $a\ge a_{Cut-off}$).  The expected 
power-law index from the radial velocity 
distribution (see Fig ~\ref{smafig}) is -0.61 \citep{cumming}, and given 
these data we can 
place a 95\% confidence limit on the upper cut-off of 75 AU.  At 68\% 
confidence, there cannot be giant planets in orbits beyond 29 AU, for this 
choice of power law index.  For this 
figure, we use the models of \citet{burrows}
  \label{smacontourfig}}
\end{figure}

\begin{figure}
\epsscale{1}
\plotone{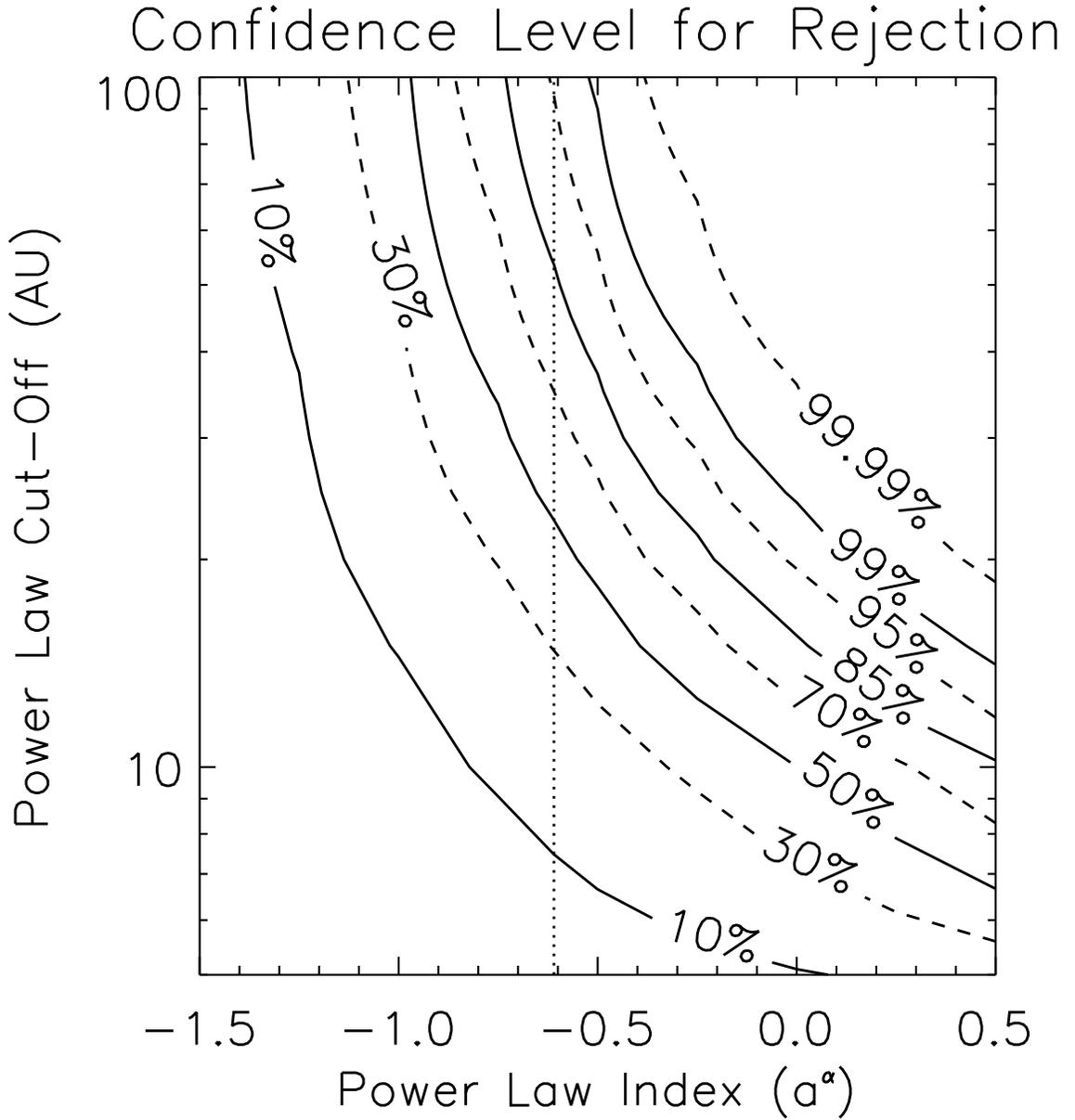}
\caption{The same as Fig.~\ref{smacontourfig}, but using the models 
of \citet{cond}.  The 95\% confidence upper cut-off for semi-major axis for 
the $\frac{dN}{da} \propto a^{\alpha}$ model now moves to 94 AU.
  \label{smacontourfiglyon}}
\end{figure}

\begin{figure}
\epsscale{1}
\plotone{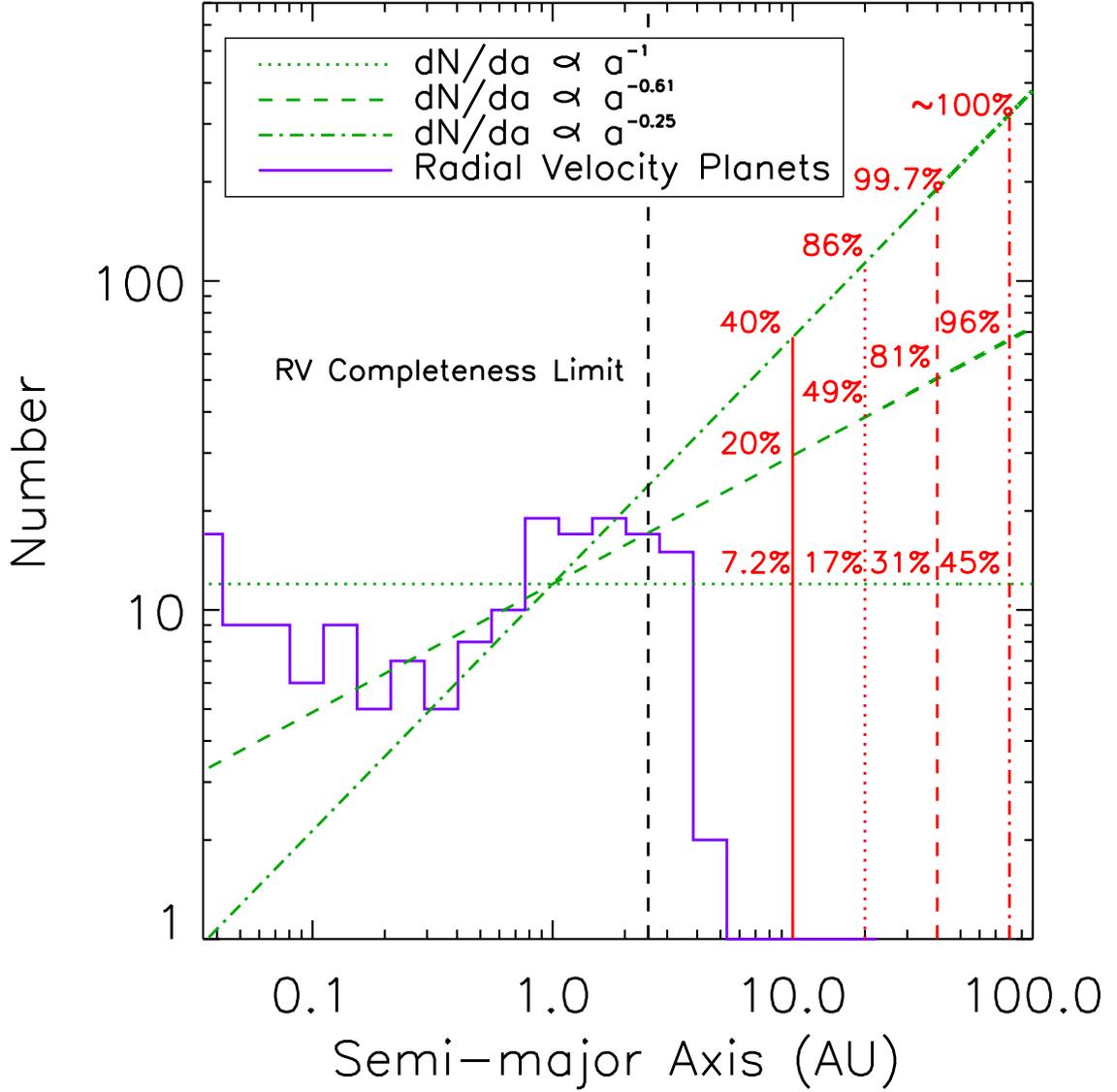}
\caption{The histogram (in blue) of the 
distribution of known extrasolar giant planets 
found with the radial velocity method, plotted against a series of power 
laws considered in Fig.~\ref{smacontourfig} and~\ref{smacontourfiglyon}.  
Since radial velocity observations are only complete to about 2.5 AU, a less 
steep drop-off of planets with semi-major axis is possible.  We give the 
confidence with which we can rule out various combinations of power law 
index and upper cut-off (the percentages in red), for indices of -1, -0.61, 
-0.25, and upper cut-offs of 10 AU, 20 AU, 40 AU, and 80 AU.  While we have 
insufficient statistics to place strong constraints on 
the power law of index -1, we can rule out the other two with increasing 
confidence as larger values of the upper limit are considered.  For example, 
a power law of the form $\frac{dN}{da} \propto a^{-0.25}$ must cut-off at 
26 AU (95\% confidence), while the most likely power law of index -0.61 
must have its cut-off at 75 AU (also at the 95\% confidence level).
  \label{dsfassumptionsfig}}
\end{figure}

In Fig.~\ref{smacontourfig} and~\ref{smacontourfiglyon}, we plot the 
confidence with which we can reject the model for various combinations of 
power law index and upper cut-off for the semi-major axis distribution.  For 
the favored model of a power law distribution given by 
$\frac{dN}{da} \propto a^{-0.61}$, we can place, at the 95\% 
confidence level, an upper-limit on the semi-major axis cut-off of 
75 AU (94 AU using the models of \citet{cond} instead of those of 
\citet{burrows}).  In other words, if the power law index has a value of 
-0.61, there can be no planets in orbits beyond \textit{a}=75 AU at the 95\% 
confidence 
level (29 AU at the 68\% confidence level).  In Fig.~\ref{dsfassumptionsfig}, 
we show how these assumptions of power law index 
compare with the distributions of known radial velocity planets, as well as 
to what confidence we can exclude various models.


\subsection{Testing Core Accretion Models}

We also consider more sophisticated models of planet populations, namely the 
core accretion models of \citet{idalin}.  Using their Fig. 12, we extract all 
the non-Hot-Jupiter giant planets, and of the 200-300 resulting planets, we 
run our Monte Carlo simulation by, for each simulated planet, randomly 
selecting one planet from this figure, adopting its values of mass and 
semi-major axis, then assigning it 
the other orbital elements as usual.  We consider each of the three cases 
modeled by \citet{idalin}.

\begin{figure}
\epsscale{1}
\plotone{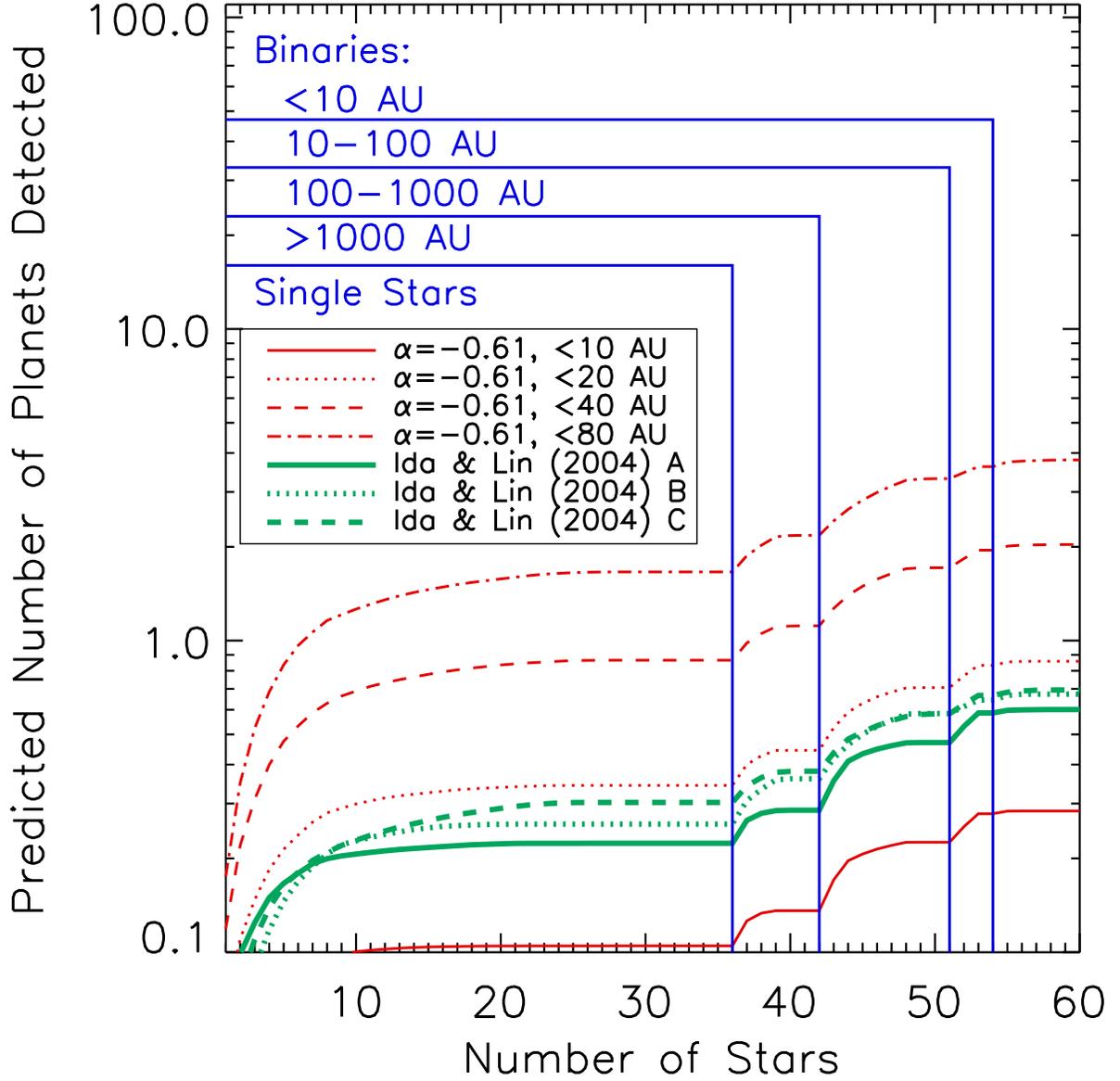}
\caption{The number of planets we would expect to detect at the end of the 
survey, 
as a function of the number of target stars observed, out of our total sample 
of 60.  Stars are divided into 
bins based on binarity, and within each bin the stars are arranged 
so that the best targets are observed first.  The first four models use power 
laws with $\frac{dN}{da} \propto a^{-0.61}$, with the upper cut-off given.  
These models can be ruled out with increasing confidence with cut-offs 
beyond 40 AU as 
increasingly close binaries are added to the sample.  Since the three 
\citet{idalin} 
models predict less than one planet from our survey, we can only place very 
limited constraints on these models at this time, namely that cases A, B, and 
C are inconsistent with our null result at the 45\%, 49\%, and 50\% 
confidence levels, respectively, if all binaries are included in our sample.
  \label{surveysizefig}}
\end{figure}

In Fig.~\ref{surveysizefig} we plot the predicted number of planets detected 
from these three distributions.  Again, the planet fraction for each curve 
is set to match the planet fraction of \citet{fv05} for planets above 1.6 
$M_{Jup}$ and within 2.5 AU.  Since the predicted total number of planets 
detected range 
between about 0.6 and 0.7 at the end of our survey, we cannot place any 
strong constraints on these models from our null result.  For the three 
cases of \citet{idalin}, A, B, and C, we can only ``rule them out'' at the 
confidence levels of 45\%, 49\%, and 50\% respectively, and again only after 
leaving all binaries in the sample.  Additionally, since we 
are considering target stars of all spectral type, we are not staying 
faithful to the original simulations of \citet{idalin}, which consider only 
solar mass host stars.  In summary, the core-accretion simulations of 
\citet{idalin} are quite consistent with our results.

\section{Discussion: Systematic Effects of Models on Results, and Other Work}\label{discussion}

We underscore the dependence of these results upon the accuracy of the 
mass-luminosity relations of \citet{burrows} and \citet{cond}.  In particular, 
these models utilize the ``Hot Start'' method for giant planet formation, at 
odds with the core accretion mechanism suggested by the planet-metallicity 
relation of \citet{fv05}.  The giant planet models of \citet{marley} 
incorporate formation by core accretion, and predict 
systematically fainter fluxes for these young planets (typically $\sim$3 
times fainter for a 30 Myr, 4 M$_{Jup}$ planet, yet the overall effect is 
difficult to predict without detailed models and spectra).  Another result 
of moving to these models, however, would be that these planets are also 
cooler, so that the SDI method (limited to objects with effective 
temperatures lower than 1400 K) will 
likely reach planets of higher masses than would be predicted by the models 
of \citet{burrows} and \citet{cond}.  

It is possible to 
envision a scenario with extrasolar planets being built by both disk 
instability (e.g. \citet{boss}) and core accretion, with the two types of 
planets segregated in orbital distance: inner planets being more common in 
orbit around metal-rich stars, consistent with core accretion, while 
outer planets (the type to which the surveys discussed here are sensitive) 
form by disk instability.  In that case, the use of the Hot Start models 
would be entirely reasonable, as these models have been shown to be mostly 
consistent with young, low-mass objects that likely form in this way 
(e.g. \citet{stassun}, 
\citet{sinfoni}).  This possibility (which we again note is pure speculation) 
endangers any conclusions drawn from Fig.~\ref{smacontourfig} 
and~\ref{smacontourfiglyon}, which assume a single, consistent population of 
planets, not allowing for the possibility of two overlapping populations 
(such as one described by broken power laws).  Our results for the upper 
limit on planet fraction would remain valid, however, since these make no 
assumptions on extrasolar planet populations beyond the eccentricity 
distribution (a minor factor) and the mass-luminosity relation.

Clearly, these constraints would be stronger with a larger sample size to 
improve our statistics.  Such an increase in sample size is hampered by the 
limited number of young, nearby stars: observing older targets tends to 
require an order of magnitude increase in number of targets so as to 
assure a similar number of detected planets.  The greatest improvement 
in these results is likely to come with more advanced planet-finding 
techniques, which increase the contrast and inner working angle to which one 
can detect planets close to their parent stars.  Two such Extreme AO systems, 
slated to come online in the 
next several years, are VLT-SPHERE and the Gemini Planet Imager (GPI).  
Surveys of a sample of young, 
nearby stars (likely very similar to the target list of this work) with these 
planet-finders should be able to greatly close the gap between the 
sensitivities to planets of direct imaging and radial velocity surveys.  

While radial velocity surveys 
continue to have great success in finding planets, the limiting factor is 
orbital time: a planet at 10 AU takes over 30 years to complete a single 
orbit, and radial velocity planets are generally not confirmable until at 
least one orbit has elapsed.  As a result, the onus on determining the 
characteristics of giant planets beyond $\sim$10 AU is largely 
upon direct imaging surveys.

A survey planned for the immediate future uses the NICI (Near Infrared 
Coronographic Imager) instrument currently being commissioned on the Gemini 
South Telescope, with plans for a 50-night survey for extrasolar giant 
planets.  It 
is hoped, of course, that these future surveys will produce actual detections, 
not just more null results, which when considered alongside the targets that 
were not found to harbor planets, should continue to constrain parameter 
space on the distribution of outer extrasolar giant planets.

Another direct imaging survey for giant planets has recently been completed, 
searching for companions to 79 young, nearby stars: the Gemini Deep 
Planet Survey \citep{gdps}.  For completeness, we run an extra set of 
simulations to compare our results to theirs.  \citet{gdps} consider the case 
of planets with masses between 0.5 and 13 M$_{Jup}$, governed by a power law 
of index -1.2 (quite similar to our value of -1.16), and with a power law of 
index -1 for semi-major axis.  They then set an upper limit on the planet 
fraction in three ranges of semi-major axis: 28\% for 10-25 AU, 13\% for 
25-50 AU, and 9.3\% for 50-200 AU, all at the 95\% confidence level, using 
the models of \citet{cond}.  Adopting these same simulation parameters, we 
find upper limits on planet fractions of 37\%, 24\%, and 28\%, respectively.  
We attribute our somewhat lower sensitivity to the increased number of 
stars in the \citet{gdps} survey, as well as their increased field of view 
(9'' compared to the 2.2'' for SDI), which makes their method better-suited 
to detecting planets at the very large orbital radii of the last two bins.  
Also, the \citet{gdps} survey was more consciously focused on closer stars: 
all 85 of their target stars are within 35 parsecs, 18 of our 60 stars are 
beyond 35 pc.  The overall results of both our work and that of 
\citet{gdps}, however, are in good 
agreement for the case of planets in shorter orbits: for example, we reach 
the same upper limits as \citet{gdps} 
reached at the 95\% confidence level, if we degrade our confidence level 
to 89\% for 10-25 AU, 80\% for 25-50 AU, and 63\% for 50-200 AU.  Hence 
the conclusions from both papers are the same: giant planets are rare 
at large separations.

We also note that the value of the planet fraction in these intervals can be 
estimated from the uniform detectability sample of \citet{fv05}, which gives 
5.5\% of stars having planets within 2.5 AU, and more massive than 1.6 
M$_{Jup}$.  When using a model of planet mass with index -1.2, and semi-major 
axis power law index -1, as above, the planet fractions for the semi-major 
axis bins 10-25 AU, 25-50 AU, and 50-200 AU become 2.1\%, 1.6\%, and 3.2\%, 
respectively.  It should be noted that the samples of \citet{fv05} and 
\citet{gdps} (as well as the one discussed in this paper, for that matter) are 
not directly comparable, as the \citet{fv05} sample 
is made up primarily of older stars 
($>$1 Gyr), and exclusively FGK spectral types, whereas the sample of 
\citet{gdps} is made up of younger stars, and contains stars of M spectral 
type.  These 
two effects push the planet fractions in opposite directions: younger stars 
are more likely to be metal-rich\footnote{Although Table 1 of \citet{gdps} 
gives 
the metallicity for most of their target stars, which give a median value of 
[Fe/H] = 0, more metal poor than the overall sample of \citet{fv05} by 
$\sim$0.1 dex, it is notoriously difficult to make accurate metallicity 
measurements of young stars.  As a result, it is likely that these 
reported metallicities are systematically lower than their actual values.}, 
and so have a higher planet 
fraction \citep{fv05}, whereas M stars are less likely to harbor giant 
planets \citep{johnson}.  Overall, then, the upper limits from both 
papers are consistent with the predictions from radial 
velocity detections, with respect to this particular model of planet 
populations.

Finally, we note that although four of our target stars do, in fact, harbor 
extrasolar planets (HIP 30034 (AB Pic) has a wide (5.5'') companion at the 
planet/brown 
dwarf boundary, while Eps Eri, HD 81040, and HD 128311 all have radial 
velocity planets), our survey can be regarded as a null result.  Even 
though these planets were orbiting our target stars, we were unable to 
detect them, as they were either outside our field of view (as with AB Pic 
B), or too faint (due to their host star's age) to be detected from our 
images, as was the case with the radial velocity planets.  The motivation 
behind our simulations is to find what population of hidden (undetected) 
planets are consistent with a lack of planet detections, and the knowledge 
of existing planets around some target stars does not change this.

\section{Conclusion}

Even without detecting extrasolar planets from our surveys, the null results 
provide a basis for setting limits on the allowable distribution of giant 
planets.  From our data, using the planet models of \citet{burrows}, 
we can exclude any model for planet 
distributions where more than 20\% of stars of all spectral types have 
planets more massive than 4$M_{Jup}$ between 20 and 100 AU, at 95\% 
confidence (this upper limit becomes 8\% of stars with such planets at the 
68\% confidence level).  If we create simple models of planet populations 
with the semi-major axis distribution governed by the 
power law $\frac{dN}{da} \propto a^{\alpha}$, and mass by 
$\frac{dN}{dM} \propto M^{-1.16}$, we can exclude giant 
planets in the case of $\alpha = 0$ beyond 18 AU, and with $\alpha = -0.5$ 
beyond 48 AU.  Using the distribution of \citet{cumming}, based on radial 
velocity observations, with $\alpha = -0.61$, there can be no giant planets 
beyond 75 AU.  All these statements are at the 95\% confidence level; for 
the 68\% confidence level, these upper limits for the outer cut-offs of 
giant planets become 12 AU, 23 AU, and 29 AU, for power law indices of 
0, -0.5, and -0.61, respectively.  With our data, the most we can say of the 
models of \citet{idalin} is that they are consistent with our observations 
at the $\sim$50\% confidence level.  We again note that these conclusions are 
highly dependent on the models of planet luminosity as a function of the 
planet's age and mass.  Additionally, we caution that since our sample 
differs from the volume-limited sample of \citet{fv05}, known correlations 
of planet fraction with stellar mass and metallicity will likely shift our 
results from the values reported here.  Nevertheless, the analysis presented 
here is an important first step in constraining the populations of 
extrasolar giant planets.

\acknowledgments

We thank the anonymous referee for many helpful comments that have improved 
the quality of this work.  We thank Eric Mamajek for a great deal of 
assistance both in selecting targets 
for the SDI survey, and determining the ages of our target stars. We thank 
Remi Soummer for the idea of presenting sensitivity to planets as a grid of 
mass and semi-major axis points, and we thank Daniel Apai for presenting the 
idea of constructing a grid of semi-major axis power law indices and 
cut-offs.  We also thank Thomas Henning and Wolfgang Brandner for their 
important work in the original data gathering, and helpful comments over the 
course of the project.  
This work makes use of data from the European Southern Observatory, 
under Program 70.C - 0777D, 70.C - 0777E, 71.C-0029A, 74.C-0548, 74.C-0549, 
and 76.C-0094.  Observations reported here were obtained at the MMT 
Observatory, a joint facility of the University of Arizona and the Smithsonian Institution.  This publication makes use of data products from the Two 
Micron All-Sky Survey, which is a joint project of the University of 
Massachusetts and the Infrared Processing and Analysis Center/California 
Institute of Technology, funded by the National Aeronautics and Space 
Administration and the National Science Foundation.  This research has made 
use of the SIMBAD database, operated at CDS, Strasburg, France.  
ELN is supported by a 
Michelson Fellowship.  LMC is supported by an NSF CAREER award and the NASA 
Origins of the Solar System program.  BAB is supported by the NASA GSRP grant 
NNG04GN95H and NASA Origins grant NNG05GL71G.

\bibliographystyle{apj}
\bibliography{apj-jour,sdi_final_enielsen01}
\clearpage

%
%
%
%

\begin{deluxetable}{lccccccccc}
\tablecolumns{9}
\tablewidth{0pc}
\tabletypesize{\tiny}
\tablecaption{Target Stars}
\tablehead{
\colhead{Target} & \colhead{RA\tablenotemark{1}} & \colhead{Dec\tablenotemark{1}} & \colhead{Distance (pc)\tablenotemark{2}} & \colhead{Sp. Type} & \colhead{Age (Myr)} & \colhead{V\tablenotemark{1}} & \colhead{H\tablenotemark{3}} & \colhead{Ks\tablenotemark{3}} & \colhead{Obs. Mode\tablenotemark{4}}}
\startdata
\citet{sdifinal} \\
\hline
HIP 1481 & 00 18 26.1 & -63 28 39.0 & 40.95 & F8/G0V & 30 & 7.46 & 6.25 & 6.15 & VLT SDI \\
HD 8558 & 01 23 21.2 & -57 28 50.7 & 49.29 & G6V & 30 & 8.54 & 6.95 & 6.85 & VLT SDI \\
HD 9054 & 01 28 08.7 & -52 38 19.2 & 37.15 & K1V & 30 & 9.35 & 6.94 & 6.83 & VLT SDI \\
HIP 9141 & 01 57 48.9 & -21 54 05.0 & 42.35 & G3/G5V & 30 & 8.11 & 6.55 & 6.47 & VLT SDI \\
BD+05 378 & 02 41 25.9 & +05 59 18.4 & 40.54 & M0 & 12 & 10.20 & 7.23 & 7.07 & VLT SDI \\
HD 17925 & 02 52 32.1 & -12 46 11.0 & 10.38 & K1V & 115 & 6.05 & 4.23 & 4.17 & VLT SDI \\
Eps Eri & 03 32 55.8 & -09 27 29.7 & 3.22 & K2V & 800 & 3.73 & 1.88 & 1.78 & VLT SDI \\
V577 Per A & 03 33 13.5 & +46 15 26.5 & 33.77 & G5IV/V & 70 & 8.35 & 6.46 & 6.37 & MMT SDI \\
GJ 174 & 04 41 18.9 & +20 54 05.4 & 13.49 & K3V & 160 & 7.98 & 5.31 & 5.15 & VLT SDI \\
GJ 182 & 04 59 34.8 & +01 47 00.7 & 26.67 & M1Ve & 12 & 10.10 & 6.45 & 6.26 & VLT SDI/Ks \\
HIP 23309 & 05 00 47.1 & -57 15 25.5 & 26.26 & M0/1 & 12 & 10.09 & 6.43 & 6.24 & VLT SDI/Ks \\
AB Dor & 05 28 44.8 & -65 26 54.9 & 14.94 & K1III & 70 & 6.93 & 4.84 & 4.69 & VLT SDI \\
GJ 207.1 & 05 33 44.8 & +01 56 43.4 & 16.82 & M2.5e & 100 & 9.50 & 7.15 & 6.86 & VLT SDI \\
UY Pic & 05 36 56.8 & -47 57 52.9 & 23.87 & K0V & 70 & 7.95 & 5.93 & 5.81 & VLT SDI \\
AO Men & 06 18 28.2 & -72 02 41.4 & 38.48 & K6/7 & 12 & 10.99 & 6.98 & 6.81 & VLT SDI/Ks \\
HIP 30030 & 06 19 08.1 & -03 26 20.0 & 52.36 & G0V & 30 & 8.00 & 6.59 & 6.55 & MMT SDI \\
HIP 30034 & 06 19 12.9 & -58 03 16.0 & 45.52 & K2V & 30 & 9.10 & 7.09 & 6.98 & VLT SDI \\
HD 45270 & 06 22 30.9 & -60 13 07.1 & 23.50 & G1V & 70 & 6.50 & 5.16 & 5.05 & VLT SDI \\
HD 48189 A & 06 38 00.4 & -61 32 00.2 & 21.67 & G1/G2V & 70 & 6.15 & 4.75 & 4.54 & VLT SDI \\
pi01 UMa & 08 39 11.7 & +65 01 15.3 & 14.27 & G1.5V & 210 & 5.63 & 4.28 & 4.17 & MMT SDI \\
HD 81040 & 09 23 47.1 & +20 21 52.0 & 32.56 & G0V & 2500 & 7.74 & 6.27 & 6.16 & MMT SDI \\
LQ Hya & 09 32 25.6 & -11 11 04.7 & 18.34 & K0V & 13 & 7.82 & 5.60 & 5.45 & MMT/VLT SDI/Ks \\
DX Leo & 09 32 43.7 & +26 59 18.7 & 17.75 & K0V & 115 & 7.01 & 5.24 & 5.12 & MMT/VLT SDI \\
HD 92945 & 10 43 28.3 & -29 03 51.4 & 21.57 & K1V & 70 & 7.76 & 5.77 & 5.66 & VLT SDI \\
GJ 417 & 11 12 32.4 & +35 48 50.7 & 21.72 & G0V & 115 & 6.41 & 5.02 & 4.96 & MMT SDI \\
TWA 14 & 11 13 26.5 & -45 23 43.0 & 46.00\tablenotemark{5} & M0 & 10 & 13.00 & 8.73 & 8.49 & VLT SDI \\
TWA 25 & 12 15 30.8 & -39 48 42.0 & 44.00\tablenotemark{5} & M0 & 10 & 11.40 & 7.50 & 7.31 & VLT SDI \\
RXJ1224.8-7503 & 12 24 47.3 & -75 03 09.4 & 24.17 & K2 & 16 & 10.51 & 7.84 & 7.71 & VLT SDI \\
HD 114613 & 13 12 03.2 & -37 48 10.9 & 20.48 & G3V & 4200 & 4.85 & 3.35 & 3.30 & VLT SDI \\
HD 128311 & 14 36 00.6 & +09 44 47.5 & 16.57 & K0 & 630 & 7.51 & 5.30 & 5.14 & MMT SDI \\
EK Dra & 14 39 00.2 & +64 17 30.0 & 33.94 & G0 & 70 & 7.60 & 6.01 & 5.91 & MMT SDI \\
HD 135363 & 15 07 56.3 & +76 12 02.7 & 29.44 & G5V & 3 & 8.72 & 6.33 & 6.19 & MMT SDI \\
KW Lup & 15 45 47.6 & -30 20 55.7 & 40.92 & K2V & 2 & 9.37 & 6.64 & 6.46 & VLT SDI \\
HD 155555 AB & 17 17 25.5 & -66 57 04.0 & 30.03 & G5IV & 12 & 7.20 & 4.91 & 4.70 & VLT SDI/Ks \\
HD 155555 C & 17 17 27.7 & -66 57 00.0 & 30.03 & M4.5 & 12 & 12.70 & 7.92 & 7.63 & VLT SDI/Ks \\
HD 166435 & 18 09 21.4 & +29 57 06.2 & 25.24 & G0 & 100 & 6.85 & 5.39 & 5.32 & MMT SDI \\
HD 172555 A & 18 45 26.9 & -64 52 16.5 & 29.23 & A5IV/V & 12 & 4.80 & 4.25 & 4.30 & VLT SDI \\
CD -64 1208 & 18 45 37.0 & -64 51 44.6 & 34.21 & K7 & 12 & 10.12 & 6.32 & 6.10 & VLT SDI/Ks \\
HD 181321 & 19 21 29.8 & -34 59 00.5 & 20.86 & G1/G2V & 160 & 6.48 & 5.05 & 4.93 & VLT SDI \\
HD 186704 & 19 45 57.3 & +04 14 54.6 & 30.26 & G0 & 200 & 7.03 & 5.62 & 5.52 & MMT SDI \\
GJ 799B & 20 41 51.1 & -32 26 09.0 & 10.22 & M4.5e & 12 & 11.00 & 0.00 & -99.00 & VLT SDI/Ks \\
GJ 799A & 20 41 51.2 & -32 26 06.6 & 10.22 & M4.5e & 12 & 10.25 & 5.20 & 4.94 & VLT SDI/Ks \\
GJ 803 & 20 45 09.5 & -31 20 27.1 & 9.94 & M0Ve & 12 & 8.81 & 4.83 & 4.53 & VLT SDI/Ks \\
HD 201091 & 21 06 53.9 & +38 44 57.9 & 3.48 & K5Ve & 2000 & 5.21 & 2.54 & 2.25 & MMT SDI \\
Eps Indi A & 22 03 21.7 & -56 47 09.5 & 3.63 & K5Ve & 1300 & 4.69 & 2.35 & 2.24 & VLT SDI \\
GJ 862 & 22 29 15.2 & -30 01 06.4 & 15.45 & K5V & 6300 & 7.65 & 5.28 & 5.11 & VLT SDI \\
HIP 112312 A & 22 44 57.8 & -33 15 01.0 & 23.61 & M4e & 12 & 12.20 & 7.15 & 6.93 & VLT SDI \\
HD 224228 & 23 56 10.7 & -39 03 08.4 & 22.08 & K3V & 70 & 8.20 & 6.01 & 5.91 & VLT SDI \\
\hline
\citet{elena} \\
\hline
HIP 2729 & 00 34 51.2 & -61 54 58 & 45.91 & K5V & 30 & 9.56 & 6.72 & 6.53 & VLT Ks \\
BD +2 1729 & 06 18 28.2 & -72 02 42 & 14.87 & K7 & 30 & 9.82 & 6.09 & 5.87 & VLT H \\
TWA 6 & 07 39 23.0 & 02 01 01 & 77.00\tablenotemark{5} & K7 & 30 & 11.62 & 8.18 & 8.04 & VLT Ks \\
BD +1 2447 & 10 18 28.8 & -31 50 02 & 7.23 & M2 & 12 & 9.63 & 5.61 & 5.31 & VLT H \\
TWA 8A & 10 28 55.5 & 00 50 28 & 21.00\tablenotemark{5} & M2 & 115 & 12.10 & 7.66 & 7.43 & VLT Ks \\
TWA 8B & 11 32 41.5 & -26 51 55 & 21.00\tablenotemark{5} & M5 & 100 & 15.20 & 9.28 & 9.01 & VLT Ks \\
TWA 9A & 11 32 41.5 & -26 51 55 & 50.33 & K5 & 800 & 11.26 & 8.03 & 7.85 & VLT Ks \\
TWA 9B & 11 48 24.2 & -37 28 49 & 50.33 & M1 & 70 & 14.10 & 9.38 & 9.15 & VLT Ks \\
SAO 252852 & 11 48 24.2 & -37 28 49 & 16.40\tablenotemark{6} & K5V & 160 & 8.47 & 5.69 & 5.51 & VLT H \\
V343 Nor & 14 42 28.1 & -64 58 43 & 39.76 & K0V & 12 & 8.14 & 5.99 & 5.85 & VLT Ks \\
PZ Tel & 15 38 57.6 & -57 42 27 & 49.65 & K0Vp & 12 & 8.42 & 6.49 & 6.37 & VLT Ks \\
BD-17 6128 & 18 53 05.9 & -50 10 50 & 47.70 & K7 & 70 & 10.60 & 7.25 & 7.04 & VLT Ks \\
\enddata
\tablenotetext{1}{from the CDS Simbad service}
\tablenotetext{2}{derived from the Hipparcos survey \citet{hip}}
\tablenotetext{3}{from the 2MASS Survey \citet{2mass}}
\tablenotetext{4}{In cases were target stars were observed by both \citet{elena} and \citet{sdifinal}, the star is listed in the \citet{sdifinal} section, with Obs. Mode given as ``VLT SDI/Ks,'' for example.}
\tablenotetext{5}{Distance from \citet{SZB03}}
\tablenotetext{6}{Distance from \citet{ZSBW01}}
\label{table1}
\end{deluxetable}

%
%
%
%

%
%
%
%
%

\begin{deluxetable}{lcccccccc}
\tablecolumns{8}
\tablewidth{0pc}
\tabletypesize{\tiny}
\tablecaption{Age Determination for Target Stars}
\tablehead{
\colhead{Target} & \colhead{Sp. Type\tablenotemark{*}} & \colhead{Li EW (mas)\tablenotemark{*}} & \colhead{Li Age (Myr)} & \colhead{R'$_{HK}$\tablenotemark{*}} & \colhead{R'$_{HK}$ Age} & \colhead{Group Membership\tablenotemark{1}} & \colhead{Group Age\tablenotemark{1}} & \colhead{Adopted Age}}
\startdata
\citet{sdifinal} \\
\hline
HIP 1481 & F8/G0V\tablenotemark{2} & 129\tablenotemark{3} & 100 & -4.360\tablenotemark{4} & 200 & Tuc/Hor & 30 & 30 \\
HD 8558 & G6V\tablenotemark{2} & 205\tablenotemark{5} & 13 &   &   & Tuc/Hor & 30 & 30 \\
HD 9054 & K1V\tablenotemark{2} & 170\tablenotemark{5} & 160 & -4.236\tablenotemark{6} & $<$100 & Tuc/Hor & 30 & 30 \\
HIP 9141 & G3/G5V\tablenotemark{7} & 181\tablenotemark{8} & 13 &   &   & Tuc/Hor & 30 & 30 \\
BD+05 378 & M0\tablenotemark{9} & 15\tablenotemark{10} &   &   &   & $\beta$ Pic & 12 & 12 \\
HD 17925 & K1V\tablenotemark{7} & 194\tablenotemark{8} & 50 & -4.357\tablenotemark{6} & 200 & Her/Lyr & 115 & 115 \\
Eps Eri & K2V\tablenotemark{11} &   &   & -4.598\tablenotemark{6} & 1300 &   &   & 800\tablenotemark{12} \\
V577 Per A & G5IV/V\tablenotemark{13} & 219\tablenotemark{13} & 3 &   &   & AB Dor & 70 & 70 \\
GJ 174 & K3V\tablenotemark{14} & 118\tablenotemark{8} & 160 & -4.066\tablenotemark{6} & $<$100 &   &   & 160 \\
GJ 182 & M1Ve\tablenotemark{10} & 280\tablenotemark{15} & 12 &   &   &   &   & 12 \\
AB Dor & K1III\tablenotemark{2} & 267\tablenotemark{8} & 10 & -3.880\tablenotemark{6} & $<$100 & AB Dor & 70 & 70 \\
GJ 207.1 & M2.5e\tablenotemark{16} &   &   &   &   &   &   & 100\tablenotemark{17} \\
HIP 23309 & M0/1\tablenotemark{18} & 294\tablenotemark{18} & 12 & -3.893\tablenotemark{6} &   & $\beta$ Pic & 12 & 12 \\
UY Pic & K0V\tablenotemark{19} & 263\tablenotemark{8} & 10 & -4.234\tablenotemark{6} & $<$100 & AB Dor & 70 & 70 \\
AO Men & K6/7\tablenotemark{18} & 357\tablenotemark{18} & 6 & -3.755\tablenotemark{6} &   & $\beta$ Pic & 12 & 12 \\
HD 45270 & G1V\tablenotemark{2} & 149\tablenotemark{5} &   &   &   & AB Dor & 70 & 70 \\
HD 48189 A & G1/G2V\tablenotemark{2} & 145\tablenotemark{8} & 25 & -4.268\tablenotemark{6} & 100 & AB Dor & 70 & 70 \\
HIP 30030 & G0V\tablenotemark{20} & 219\tablenotemark{8} & 2 &   &   & Tuc/Hor & 30 & 30 \\
HIP 30034 & K2V\tablenotemark{2} &   &   &   &   & Tuc/Hor & 30 & 30 \\
pi01 UMa & G1.5V\tablenotemark{21} & 135\tablenotemark{8} & 100 & -4.400\tablenotemark{22} & 320 &   &   & 210 \\
DX Leo & K0V\tablenotemark{21} & 180\tablenotemark{8} & 100 & -4.234\tablenotemark{6} & $<$100 & Her/Lyr & 115 & 115 \\
HD 81040 & G0V\tablenotemark{21} & 24\tablenotemark{23} & 2500 &   &   &   &   & 2500 \\
LQ Hya & K0V\tablenotemark{21} & 247\tablenotemark{8} & 13 &   &   &   &   & 13 \\
HD 92945 & K1V\tablenotemark{21} & 138\tablenotemark{8} & 160 & -4.393\tablenotemark{6} & 320 & AB Dor & 70 & 70 \\
GJ 417 & G0V\tablenotemark{24} & 76\tablenotemark{25} & 250 & -4.368\tablenotemark{26} & 250 & Her/Lyr & 115 & 115 \\
TWA 14 & M0\tablenotemark{27} & 600\tablenotemark{27} & 8 &   &   & TW Hya & 10 & 10 \\
RXJ1224.8-7503 & K2\tablenotemark{28} & 250\tablenotemark{28} & 16 &   &   &   &   & 16 \\
TWA 25 & M0\tablenotemark{9} & 494\tablenotemark{29} & 10 &   &   & TW Hya & 10 & 10 \\
HD 114613 & G3V\tablenotemark{30} & 100\tablenotemark{31} & 400 & -5.118\tablenotemark{6} & 7900 &   &   & 4200 \\
EK Dra & G0\tablenotemark{32} & 212\tablenotemark{8} & 2 & -4.180\tablenotemark{22} & $<$100 & AB Dor & 70 & 70 \\
HD 128311 & K0\tablenotemark{21} &   &   & -4.489\tablenotemark{26} & 630 &   &   & 630 \\
HD 135363 & G5V\tablenotemark{21} & 220\tablenotemark{8} & 3 &   &   &   &   & 3 \\
KW Lup & K2V\tablenotemark{30} & 430\tablenotemark{33} & 2 &   &   &   &   & 2 \\
HD 155555 AB & G5IV\tablenotemark{18} & 205\tablenotemark{8} & 6 & -3.965\tablenotemark{6} & $<$100 & $\beta$ Pic & 12 & 12 \\
HD 155555 C & M4.5\tablenotemark{18} &   &   &   &   & $\beta$ Pic & 12 & 12 \\
CD -64 1208 & K7\tablenotemark{18} & 580\tablenotemark{18} & 5 &   &   & $\beta$ Pic & 12 & 12 \\
HD 166435 & G0\tablenotemark{34} &   &   & -4.270\tablenotemark{22} & 100 &   &   & 100 \\
HD 172555 A & A5IV/V\tablenotemark{2} &   &   &   &   & $\beta$ Pic & 12 & 12 \\
HD 181321 & G1/G2V\tablenotemark{30} & 131\tablenotemark{8} & 79 & -4.372\tablenotemark{6} & 250 &   &   & 160 \\
HD 186704 & G0\tablenotemark{35} &   &   & -4.350\tablenotemark{22} & 200 &   &   & 200 \\
GJ 799A & M4.5e\tablenotemark{16} &   &   &   &   & $\beta$ Pic & 12 & 12 \\
GJ 799B & M4.5e\tablenotemark{16} &   &   &   &   & $\beta$ Pic & 12 & 12 \\
GJ 803 & M0Ve\tablenotemark{16} & 51\tablenotemark{8} & 30 &   &   & $\beta$ Pic & 12 & 12 \\
HD 201091 & K5Ve\tablenotemark{16} &   &   & -4.704\tablenotemark{6} & 2000\tablenotemark{+} &   &   & 2000 \\
Eps Indi A & K5Ve\tablenotemark{16} &   &   & -4.851\tablenotemark{6} & 4000 &   &   & 1300\tablenotemark{36} \\
GJ 862 & K5V\tablenotemark{16} & 5\tablenotemark{15} &   & -4.983\tablenotemark{6} & 6300\tablenotemark{+} &   &   & 6300 \\
HIP 112312 A & M4e\tablenotemark{9} &   &   &   &   & $\beta$ Pic & 12 & 12 \\
HD 224228 & K3V\tablenotemark{30} & 53\tablenotemark{8} & 630 & -4.468\tablenotemark{6} & 500 & AB Dor & 70 & 70 \\
\hline
\citet{elena} \\
\hline
HIP 2729 & K5V\tablenotemark{2} &   &   &   &   & Tuc/Hor & 30 & 30 \\
BD +2 1729 & K7\tablenotemark{21} &   &   &   &   & Her/Lyr & 115 & 115 \\
TWA 6 & K7\tablenotemark{37} & 560\tablenotemark{37} & 3 &   &   & TW Hya & 10 & 10 \\
BD +1 2447 & M2\tablenotemark{38} &   &   &   &   & TW Hya & 150 & 150 \\
TWA 8A & M2\tablenotemark{37} & 530\tablenotemark{37} & 3 &   &   & TW Hya & 10 & 10 \\
TWA 8B & M5\tablenotemark{37} & 560\tablenotemark{37} & 3 &   &   & TW Hya & 10 & 10 \\
TWA 9A & K5\tablenotemark{37} & 460\tablenotemark{37} & 3 &   &   & TW Hya & 10 & 10 \\
TWA 9B & M1\tablenotemark{37} & 480\tablenotemark{37} & 3 &   &   & TW Hya & 10 & 10 \\
SAO 252852 & K5V\tablenotemark{39} &   &   &   &   & Her/Lyr & 115 & 115 \\
V343 Nor & K0V\tablenotemark{2} & 300\tablenotemark{31} & 5 &   &   & $\beta$ Pic & 12 & 12 \\
PZ Tel & K0Vp\tablenotemark{19} & 267\tablenotemark{40} & 20 &   &   & $\beta$ Pic & 12 & 12 \\
BD-17 6128 & K7\tablenotemark{41} & 400\tablenotemark{42} & 3 &   &   & $\beta$ Pic & 12 & 12 \\
\enddata
\tablenotetext{1}{Group Membership for TWA, $\beta$ Pic, Tuc/Hor, and AB Dor from \citet{ZS04}, Her/Lyr from \citet{LMCF06}.  Group Ages from \citet{ZS04} (TWA, $\beta$ Pic, and Tuc/Hor), \citet{abdorme} (AB Dor), and \citet{LMCF06} (Her/Lyr)}
\tablenotetext{*}{Measurement References: 2: \citet{HC75},  3: \citet{WCMM05},  4: \citet{Henry96},  5: \citet{TDQDJ00},  6: \citet{Gray06},  7: \citet{HS88},  8: \citet{WSH03},  9: \citet{ZS04},  10: \citet{FBMS95},  11: \citet{CHW67},  12: \citet{benedict},  13: \citet{CM02},  14: \citet{LP60},  15: \citet{FMS97},  16: \citet{GJ91},  17: \citet{LBSKW05},  18: \citet{ZSBW01},  19: \citet{H78},  20: \citet{CPKR95},  21: \citet{MLGFDC01},  22: \citet{Wright04},  23: \citet{SUZT06},  24: \citet{B51},  25: \citet{GHH00},  26: \citet{Gray03},  27: \citet{ZWSB01},  28: \citet{AKSCWM95},  29: \citet{SZB03},  30: \citet{H82},  31: \citet{RGP93},  32: \citet{GJ79},  33: \citet{NB98},  34: \citet{HDC},  35: \citet{A85},  36: \citet{lachaume},  37: \citet{WZPPWSM99},  38: \citet{VJMW46},  39: \citet{E61},  40: \citet{SKH98},  41: \citet{NKAVB95},  42: \citet{MDCKDS95}}
\tablenotetext{+}{In general, we have only determined Ca R'$_{HK}$ ages for stars with spectral types K1 or earlier, but in the case of these two K5 stars, we have only the R'$_{HK}$ measurement on which to rely for age determination.  The calibration of Mt. Wilson  S-index to R'$_{HK}$ for K5 stars (B-V $\sim$ 1.1 mag) has not been well-defined (\citet{noyes}; specifically the photospheric subtraction), and hence applying a R'$_{HK}$ vs. age relation for K5 stars is unlikely to yield useful ages.  Although we adopt specific values for the ages of these stars, it would be more accurate to state simply that these stars have ages $>$1 Gyr.  As a result, almost all simulated planets are too faint to detect around these stars, so the precise error in the age does not significantly affect our final results.}
\label{table2}
\end{deluxetable}

%
%
%
%
\begin{deluxetable}{lcccc}
\tablecolumns{4}
\tablewidth{0pc}
\tabletypesize{\tiny}
\tablecaption{Binaries}
\tablehead{
\colhead{Target} & \colhead{Sep (``)} & \colhead{Sep. (AU)} & \colhead{Reference} & \colhead{Companion Type}}
\startdata
\citet{sdifinal} \\
\hline
HIP 9141 & 0.15 & 6.38 & \citet{sdifinal} & \\
V577 Per A & 7 & 230 & \citet{PABBB93} & M0 \\
AB Dor & 9 (Ba/Bb) & 134 (Ba/Bb) & \citet{abdor} & Binary M stars\\
AB Dor & 0.15 (C) & 2.24 (C) & \citet{abdor} & Very low-mass M Star\\
HIP 30034 & 5.5 & 250 & \citet{abpic} & Planet/Brown Dwarf \\
HD 48189 A & 0.76 (B) & 16.5 & \citet{FM00} & K star \\
HD 48189 A & 0.14 & 3.03 & \citet{sdifinal} & \\
DX Leo & 65 & 1200 & \citet{LBSKW05} & M5.5 \\
EK Dra & SB & SB & \citet{MH06} & M2\\
HD 135363 & 0.26 & 7.65 & \citet{sdifinal} & \\
HD 155555 AB & SB (AB) & SB (AB) & \citet{BEL67} & G5 and K0 SB \\
HD 155555 AB & 18 (C) & 1060 (C) & \citet{ZSBW01} & Target Star 155555 C, M4.5 \\
HD 172555 A & 71 & 2100 & \citet{SD93} & Target Star CD -64 1208, K7 \\
HD 186704 & 13 & 380 & \citet{A32} & \\
GJ 799A & 3.6 & 36 & \citet{W52} & Target Star GJ 799B, M4.5 \\
HD 201091 & 16 & 55 & \citet{B50} & K5 \\
Eps Indi A & 400 & 1500 & \citet{epsindi} & Binary Brown Dwarf \\
HIP 112312 & 100 & 2400 & \citet{SBZ02} & M4.5 \\

\hline
\citet{elena} \\
\hline

TWA 8A & 13 & 270 & \citet{JHFF99} & Target Star TWA 8B, M5 \\
TWA 9A & 9 & 576 & \citet{JHFF99} & Target Star TWA 9B, M1 \\
SAO 252852 & 15.7 & 260 & \citet{poveda} & HD 128898, Ap \\
V343 Nor & 10 & 432 & \citet{SZB03} & M4.5 \\
BD-17 6128 & 2 & 100 & \citet{NGMGE02} & M2\\

\enddata
\label{table3}
\end{deluxetable}

\clearpage




\end{document}